\documentclass[aps,pre,twocolumn,superscriptaddress,floatfix,showpacs]{revtex4}

\usepackage{amssymb,amsmath,graphicx}

\begin{document}

\title{Amplitude death in oscillator networks with variable-delay coupling}

\author{Aleksandar Gjurchinovski}

\email{agjurcin@pmf.ukim.mk}

\affiliation{Institute of Physics, Faculty of Natural Sciences and Mathematics, Sts.\ Cyril and Methodius University,
P.\ O.\ Box 162, 1000 Skopje, Macedonia}

\author{Anna Zakharova}

\email{anna.zakharova@tu-berlin.de}

\affiliation{Institut f\"ur Theoretische Physik, Technische Universit\"at Berlin, 10623 Berlin, Germany}

\author{Eckehard Sch\"oll}

\email{schoell@physik.tu-berlin.de}

\affiliation{Institut f\"ur Theoretische Physik, Technische Universit\"at Berlin, 10623 Berlin, Germany}

\begin{abstract}
We study the conditions of amplitude death in a network of delay-coupled limit cycle oscillators by including time-varying delay in the coupling and self-feedback. By generalizing the master stability function formalism to include variable-delay connections with high-frequency delay modulations (i.e., the distributed-delay limit), we analyze the regimes of amplitude death in a ring network of Stuart-Landau oscillators and demonstrate the superiority of the proposed method with respect to the constant delay case. 
The possibility of stabilizing the steady state is restricted by the odd-number property of the local node dynamics independently of the network topology and the coupling parameters.
\end{abstract}

\pacs{05.45.Xt, 02.30.Ks}

\date{\today}

\maketitle

\section{Introduction}

Control of collective dynamics of populations of coupled nonlinear oscillators has been a subject of intensive research from both theoretical and practical aspects in the past decades \cite{KUR84,PIK01,SCH07,ATA10,SUN13}. Complex oscillator systems whose elements are in mutual interaction in a prescribed manner within a specific network topology give rise to a rich variety of emergent spatio-temporal patterns \cite{BAR08,ARE08,BOC06,NEW06}. Depending on the parameters of the coupling between the individual oscillators, the network dynamics may exhibit different kinds of synchronized behavior \cite{KIN09,FLU10,CHO10,HEI11,DAH12}, or may show oscillation suppression (quenching) towards stabilizing a collective steady state \cite{TUR52,PRI68,BAR85,ARO90,ERM90,MIR90,RES11,RES12}. In the latter situation, two structurally different types of oscillation quenching can be distinguished, depending on whether the stabilization leads to a homogeneous (amplitude death) or inhomogeneous (oscillation death) steady state \cite{SAX12,KOS13}. Oscillation death is typical for biological systems \cite{KUZ04,TSA06}, where it has been recognized as a possible mechanism of pattern formation and cellular differentiation \cite{KOS10a,SUZ11,ZAK13a}. Amplitude death, on the other hand, is an important practical mechanism in stabilizing homogeneous steady states, which is relevant, for example, in regulating the fluctuations of the output in coupled laser systems \cite{KIM05,GAN11,SOR13}, or in suppression of the pathological rhythms in an ensemble of coupled neurons related to some specific neuronal disorders \cite{ERM90,ROS04a,ROS04b,POP05,SUA09,SCH09,JIR10}, e.g. epilepsy or Parkinson's disease. Oscillation quenching has been observed in various real experiments, such as chemical \cite{YOS97,CRO89,TOI08} and electrochemical oscillators \cite{ZHA04}, electronic oscillators \cite{OZD04,LIU05,HEI10}, coupled laser systems \cite{KIM05,BIE93,PYR00,VIC06}, climate models \cite{GAL01}, ecological models \cite{EUR05}, epidemical models \cite{
NEA07}, neural networks \cite{HE01,OMI07}, etc., suggesting a potential importance in many practical applications.  

The relevance of controlling the collective steady state in oscillator networks by amplitude death mechanisms is particularly prominent in those situations when the internal parameters of the individual network units cannot be accessed or changed by external means. In this case, the control of the steady state may be achieved by appropriately modifying the form of interactions between the coupled network elements. There are several known scenarios that can lead to amplitude death in this case, the most important one being the parameter mismatch (e.g., frequency mismatch) between the oscillators \cite{BAR85,ERM90,ZHA04}, coupling through dissimilar variables (conjugate coupling) \cite{KAR07,DAS10}, or introducing delays in the interactions between different units \cite{RAM98,RAM99,KON04,SON11,ZOU13}. In the latter case, it has been shown that amplitude death occurs for a much larger set of coupling parameters when the time delay in the connections is distributed over an interval with respect to the case when the delay is fixed at a certain value \cite{ATA03,KYR11,KYR13}. Taking into account the propagation delays when modeling real systems inevitably leads to a more realistic situation than considering instantaneous connection only, since the propagation time of the signals between different units of the network system is important, and often distributed or time-varying in an interval. In this context, it has been shown for single systems that using a time-varying delayed feedback with either deterministic or stochastic variation of the delay time can considerably enlarge the stability region of the unstable steady states, thus making the fixed point stabilization more robust with respect to changes in the control parameters \cite{GJU08,JUE12,GJU13}. High-frequency modulation of the delay time is effectively equivalent to a distributed delayed feedback with a related delay distribution in the interval of delay variation \cite{MIC05}, thus rendering the time-varying delay method an efficient experimental way for 
realizing distributed delays with desired distribution kernels \cite{JUE12}. 

In this paper we propose a method for controlling amplitude death in networks of delay-coupled limit cycle oscillators with a general variable-delay interconnection between the oscillators. In the analysis, we employ the formalism of the master stability function \cite{PEC98}, commonly used to analyze the synchronous dynamics of complex networks. In addition to the interconnection topology of the coupled system, we also consider the influence of a variable-delay self-feedback either at each oscillator, or at a single oscillator only. 
In networks of delay-coupled laser systems \cite{SOR13,FLU11}, the coupling delay associated with the time of propagation of light between the units can be modulated by changing the distance between the units in a periodic fashion. Equivalently, a modulated self-feedback delay can be accomplished by periodically changing some characteristic lengths in the individual units, e.g., the width of the external cavity resonator, by applying a periodic voltage (piezoelectric effect).  Suppressing optical intensity pulsations (relaxation oscillations) and stabilizing the cw (continuous wave) emission steady state is often desirable in such realization \cite{SCHI06,FLU07,DAH08b}. On the other hand, suppression of synchronized oscillations of neural networks is of practical importance for eventual treatment of various types of pathological rhythmic neural activities, such as Parkinson's disease, tremor, or epilepsy, by quenching the undesired oscillations by deep-brain stimulation \cite{POP05,TUK07}. In this case, the delayed self-feedback can be realized by applying an electrical stimulus via implanted microelectrodes in the brain, and the signal delay can be modulated appropriately, e.g. by using a digital delay line. Furtheron, in technological networks \cite{ARE08}, like power grids, sensor networks, communication networks, etc., high-frequency modulation of the delay times might be used at purpose to enhance the stability range of the steady state.

The organization of the paper is as follows. The stability analysis of the model is performed in Sec. II both for the general case and for a high-frequency modulation of the delays. In the latter case, the model system is approximated by a related distributed-delay system \cite{GJU13}, thus enabling a stability analysis via the master stability function formalism. Two limitations of achieving amplitude death are pointed out, which are particularly relevant in practical implementations of the method. The results of the stability analysis are applied in Sec. III to investigate the regimes of amplitude death in various control parameter spaces for a regular ring network of delay-coupled Stuart-Landau oscillators. The analysis is performed (i) for the case without self-feedback, (ii) when the self-feedback is applied at each node, and (iii) when the self-feedback is at a single node only. Each of these cases is relevant in experimental realizations of both natural and man-made networks. We conclude in Sec. IV.

\section{Linear stability analysis}

\subsection{General master stability equation}

We consider a network consisting of $N$ nodes populated with identical oscillators with a variable-delay coupling between different oscillators and variable-delay self-feedback. The model equations are given by:
\begin{align}
\dot{\mathbf{x}_j}=\mathbf{f}(\mathbf{x}_j)&+\sigma_1\sum_{n=1}^N a_{jn}\widehat{\mathbf{H}}_1\left(\mathbf{x}_n(t-\tau_1(t))-\mathbf{x}_j(t)\right)\nonumber\\
&+\sigma_2\widehat{\mathbf{H}}_2\left(\mathbf{x}_j(t-\tau_2(t))-\mathbf{x}_j(t)\right),
\label{system}
\end{align}
where $\mathbf{x}_j\in\mathbb{R}^d$ is a $d$-dimensional state vector of the $j$-th oscillator at vertex $j$ ($j=1,2\dots N$). The intrinsic dynamics of each oscillator is specified by a $d$-dimensional non-linear vector function $\mathbf{f}:\mathbb{R}^d\rightarrow\mathbb{R}^d$.
The $N \times N$ matrix $\widehat{\mathbf{A}}$ is the adjacency matrix of the network excluding the self-feedback, i.e., it determines only the connection topology between adjacent nodes, and $\tau_1(t)$ is the time-varying coupling delay. The element $a_{jn}$ is unity if node $n$ is connected to node $j$, and is zero otherwise, i.e. the matrix $\widehat{\mathbf{A}}$ is a Boolean matrix with zero entries along the diagonal ($a_{jj}=0$). The self-feedback terms are taken in the form of a linear Pyragas-type control \cite{PYR92} with a variable time delay $\tau_2(t)$.
The parameters $\sigma_1$ and $\sigma_{2}$ are the strengths of coupling and self-feedback, respectively. The coupling scheme between different components on two adjacent nodes is described by a $d\times d$ matrix $\widehat{\mathbf{H}}_1$, and $\widehat{\mathbf{H}}_2$ is the corresponding $d\times d$ matrix of the self-feedback. We assume that $\sigma_1$, $\sigma_2$, $\widehat{\mathbf{H}}_1$ and $\widehat{\mathbf{H}}_2$ are the same at each node, but they can differ in general case.

In this paper, we consider networks in which each oscillator element has the same node degree in the interconnection topology, i.e.,  the interconnection adjacency matrix $\widehat{\mathbf{A}}$ has a constant row sum $\mu$:
\begin{equation}
\sum_{n=1}^N a_{jn}=\mu,
\label{constrowsum}
\end{equation}
which is independent of $j$. 
Below we show that this condition allows for a substantial simplification in the stability analysis by using the formalism of the master stability function \cite{PEC98}. The time-delay functions $\tau_1(t)$ and $\tau_2(t)$ are modulated around average delay values $\tau_{01}$ and $\tau_{02}$. We consider periodic deterministic modulations in the form
\begin{align}
\tau_1(t)&=\tau_{01}+\varepsilon_1 \Phi_1(\varpi_1 t),\\
\tau_2(t)&=\tau_{02}+\varepsilon_2 \Phi_2(\varpi_2 t),
\label{delayvar}
\end{align}
where $\Phi_{1,2}:\mathbb{R}\rightarrow[-1,1]$ are $2\pi$-periodic functions with zero mean, and $\varepsilon_{1,2}$ and $\varpi_{1,2}$ are the amplitudes and the angular frequencies of the corresponding delay modulations, respectively. 

In the following, we investigate the stability of the steady state solution of the oscillator network.
Collecting the dynamical variables $\mathbf{x}_j$ of all $N$ oscillators into a single $Nd$-dimensional state vector $\mathbf{X}=(\mathbf{x}_1,\mathbf{x}_2,\dots,\mathbf{x}_N)^{T}$,
the collective dynamics of the network is given by:
\begin{align}
\dot{\mathbf{X}}=&\mathbf{F}\left(\mathbf{X}(t)\right)-\sigma_1\mu\left(\widehat{\mathbf{I}}_{N}\otimes\widehat{\mathbf{H}}_1\right)\mathbf{X}(t)\nonumber\\ &+\sigma_1\left(\widehat{\mathbf{A}}\otimes\widehat{\mathbf{H}}_1\right)\mathbf{X}(t-\tau_1(t))\nonumber\\
&+\sigma_2\left(\widehat{\mathbf{I}}_{N}\otimes\widehat{\mathbf{H}}_2\right)\left(\mathbf{X}(t-\tau_2(t))-\mathbf{X}(t)\right)
\label{collective}
\end{align}
where $\widehat{\mathbf{I}}_N$ is $N\times N$ identity matrix, and $\mathbf{F}(\mathbf{X}(t))=\left(\mathbf{f}(\mathbf{x}_1),\mathbf{f}(\mathbf{x}_2),\dots,\mathbf{f}(\mathbf{x}_N)\right)^{T}$ is an $N d$-dimensional vector field. 
We assume that the system (\ref{collective}) has a symmetric (homogeneous) fixed point  $\mathbf{X}^*=(\mathbf{x}^*,\mathbf{x}^*,\dots,\mathbf{x}^*)^{T}$. The components $\mathbf{x}_j(t)=\mathbf{x}^*$ of the fixed point are the same at each node  $j=1,2\dots N$, and from Eq. (\ref{system}) we have $\mathbf{f}(\mathbf{x}^*)=\mathbf{0}$. This means that each coupled oscillator has an identical steady state as in the absence of any connection between nodes. The existence of such a homogeneous steady state is a consequence of the chosen form of interaction between the nodes, making the position of the symmetric fixed point of the network independent of the connection topology. In this case, a constant row-sum of the interconnection adjacency matrix is not a requirement for the existence of a collective symmetric steady state solution.   

We assume instability of the collective fixed point $\mathbf{X}^*$ in the uncoupled oscillatory regime, and investigate the influence of the variable-delay coupling and the network topology on the stability.
Considering a small deviation from the fixed point $\delta\mathbf{X}(t)=\mathbf{X}(t)-\mathbf{X}^*$, we arrive at a variational equation:
\begin{align}
\delta\dot{\mathbf{X}}(t)=&\left(\widehat{\mathbf{I}}_N\otimes\widehat{\mathbf{J}}\right)\delta\mathbf{X}(t)
-\sigma_1\mu\left(\widehat{\mathbf{I}}_{N}\otimes\widehat{\mathbf{H}}_1\right)\delta\mathbf{X}(t)\nonumber\\ &+\sigma_1\left(\widehat{\mathbf{A}}\otimes\widehat{\mathbf{H}}_1\right)\delta\mathbf{X}(t-\tau_1(t))\nonumber\\
&+\sigma_2\left(\widehat{\mathbf{I}}_{N}\otimes\widehat{\mathbf{H}}_2\right)\left(\delta\mathbf{X}(t-\tau_2(t))-\delta\mathbf{X}(t)\right)
\label{varequat}
\end{align}
where $\widehat{\mathbf{J}}=D[\mathbf{f}(\mathbf{x}^*)]$ is the $d\times d$ Jacobian matrix of a single oscillator without coupling, calculated at the fixed point $\mathbf{x}^*$ of a single node.

To achieve a block-diagonalization of the term involving the adjacency matrix $\widehat{\mathbf{A}}$, assuming that $\widehat{\mathbf{A}}$ can be diagonalized, we multiply Eq. (\ref{varequat}) with $\widehat{\mathbf{S}}\otimes\widehat{\mathbf{I}}_d$ from left, where  $\widehat{\mathbf{I}}_d$ is $d\times d$ identity matrix, and $\widehat{\mathbf{S}}$ is a $N\times N$ matrix that diagonalizes $\widehat{\mathbf{A}}$ ($\widehat{\mathbf{A}}_{\mathrm{diag}}=\widehat{\mathbf{S}}\widehat{\mathbf{A}}\widehat{\mathbf{S}}^{-1}$). By using the properties of the Kronecker product, we obtain:
\begin{align}
\delta\dot{\mathbf{\widetilde{X}}}(t)=&\widehat{\mathbf{I}}_N\otimes\left(\widehat{\mathbf{J}}-\sigma_1\mu\widehat{\mathbf{H}}_1-\sigma_2\widehat{\mathbf{H}}_2\right)\delta\mathbf{\widetilde{X}}(t)\nonumber\\
&+\sigma_1\left(\widehat{\mathbf{A}}_{\mathrm{diag}}\otimes\widehat{\mathbf{H}}_1\right)\delta\mathbf{\widetilde{X}}(t-\tau_1(t))\nonumber\\
&+\sigma_2\left(\widehat{\mathbf{I}}_N\otimes\widehat{\mathbf{H}}_2\right)\delta\mathbf{\widetilde{X}}(t-\tau_2(t)),
\label{vareq}
\end{align}
where $\mathbf{\widetilde{X}}(t)=\left(\widehat{\mathbf{S}}\otimes\widehat{\mathbf{I}}_d\right)\mathbf{X}(t)$, and 
\begin{equation}
\widehat{\mathbf{A}}_{\mathrm{diag}}=\mathrm{diag}(\mu,\nu_2,\nu_3\dots,\nu_{N}) 
\end{equation}
is the diagonalized adjacency matrix, containing its eigenvalues $\nu_j$ $(j=1,2\dots N)$ along the diagonal. 
Here, $\nu_1=\mu$ is the row-sum of $\widehat{\mathbf{A}}$, which is always an eigenvalue of $\widehat{\mathbf{A}}$ corresponding to the perturbation direction along the $N$-dimensional eigenvector $(1,1,\dots,1)^T$. When investigating the (transverse) stability of the synchronous periodic solution, the eigenvalue $\nu_1=\mu$ is not involved, since the master stability equation for $\nu_1=\mu$ (\textit{longitudinal eigenvalue}) corresponds to the variational equation on the synchronization manifold. However, for the stability of the collective fixed point $\mathbf{X}^*$ of the network, each perturbation direction from the fixed point matters for the stability, and all the eigenvalues of the adjacency matrix are equally involved in determining the stability. 

Since the resulting variational equation (\ref{vareq}) has a block structure with $N$ independent blocks ($m=1,\dots,N$), each block can be considered separately in the stability analysis:
\begin{align}
\delta\dot{\mathbf{\widetilde{x}}}&_m(t)=\left(\widehat{\mathbf{J}}-\sigma_1\mu\widehat{\mathbf{H}}_1-\sigma_2\widehat{\mathbf{H}}_2\right)
\delta\mathbf{\widetilde{x}}_m(t)\nonumber\\
&+\sigma_1\nu_m\widehat{\mathbf{H}}_1\,\delta\mathbf{\widetilde{x}}_m(t-\tau_1(t))
+\sigma_2\widehat{\mathbf{H}}_2\,\delta\mathbf{\widetilde{x}}_m(t-\tau_2(t)),
\end{align}
where $\delta\mathbf{\widetilde{x}}_m(t)$ is a $d$-dimensional state vector of the $m$-th node in the new coordinates. Hence, the master stability equation of the network system is:
\begin{align}
\delta\dot{\mathbf{\widetilde{x}}}(t)&=\left(\widehat{\mathbf{J}}-\sigma_1\mu\widehat{\mathbf{H}}_1-\sigma_2\widehat{\mathbf{H}}_2\right)
\delta\mathbf{\widetilde{x}}(t)\nonumber\\
&+\sigma_1\nu\widehat{\mathbf{H}}_1\,\delta\mathbf{\widetilde{x}}(t-\tau_1(t))
+\sigma_2\widehat{\mathbf{H}}_2\,\delta\mathbf{\widetilde{x}}(t-\tau_2(t)),
\label{mse}
\end{align}
where $\nu\in\mathbb{C}$.
The fixed point $\mathbf{X}^*$ is locally asymptotically stable if and only if the perturbation $\delta\mathbf{\widetilde{x}}$ asymptotically tends toward zero for all eigenvalues $\nu_m$ of the adjacency matrix. Equivalently,  $\mathbf{X}^*$ is locally asymptotically stable if and only if the maximum real part of the characteristic exponents $\Lambda_l(\nu_m)$ ($l=1,\dots d$) arising from the master stability equation (\ref{mse}) is negative for all $\nu_m$. The function $\mathrm{max}\mathrm{Re}\{\Lambda(\nu)\}$ is the master stability function of the network. In general, the master stability function can be obtained numerically by simulating the master stability equation (\ref{mse}) for different values of $\nu \in\mathbb{C}$. If this gives rise to a region in the $(\mathrm{Re}[\nu],\mathrm{Im}[\nu])$ plane where $\mathrm{max}\mathrm{Re}\{\Lambda(\nu)\}<0$, and if all the eigenvalues of the adjacency matrix are located inside this stability region, the fixed point is locally asymptotically stable. It is unstable if at least one eigenvalue lies outside this region.

\subsection{High-frequency delay modulation}

An analytical investigation of the master stability function of the collective fixed point $\mathbf{X}^*$ of the network is possible if the frequencies $\varpi_1$ and $\varpi_2$ of the delay variations are large compared to the intrinsic eigenfrequencies of the system dynamics. In this case, the coupled oscillator system is in the regime of a distributed-delay limit \cite{MIC05,GJU13}, and the time-varying delay can be approximately replaced with a distributed delay with probability density function $\rho_{1,2}$, in which case the master stability equation reads:
\begin{align}
\delta\dot{\mathbf{\widetilde{x}}}(t)=&\left(\widehat{\mathbf{J}}-\sigma_1\mu\widehat{\mathbf{H}}_1-\sigma_2\widehat{\mathbf{H}}_2\right)
\delta\mathbf{\widetilde{x}}(t)\nonumber\\
&+\sigma_1\nu\widehat{\mathbf{H}}_1\,\int_{0}^{\infty}\rho_1(\theta)\delta\mathbf{\widetilde{x}}(t-\theta)\,d\theta\nonumber\\
&+\sigma_2\widehat{\mathbf{H}}_2\,\int_{0}^{\infty}\rho_2(\theta)\delta\mathbf{\widetilde{x}}(t-\theta)\,d\theta.
\end{align}
The distributed-delay kernels $\rho_1(\theta)$ and $\rho_2(\theta)$ are defined such that $\rho_{1,2}(\theta)d\theta$ gives the fraction of time for which $\tau_{1,2}(t)$  lies between $\theta$ and $\theta+d\theta$, satisfying $\rho_{1,2}(\theta)\geq 0$ and the probability normalization conditions
\begin{equation}
\int_0^\infty\rho_{1}(\theta)\;d\theta=1,\hspace{0.5cm} \int_0^\infty\rho_{2}(\theta)\;d\theta=1.
\end{equation} 
When $\rho_{1,2}(\theta)=\delta(0)$, where $\delta(\cdot)$ is the Dirac delta function, the interaction becomes instantaneous, without delay, and the choice $\rho_{1,2}(\theta)=\delta(\tau')$ results in a discrete delay interaction $\tau_{1,2}(t)=\tau'=const$. Various combinations for the delay kernels are possible, e.g., a constant delay inter-node connection and a variable-delay self-feedback, or vice versa.

\begin{table*}
\caption{Various delay modulation functions $\Phi$, and corresponding distributed-delay kernels $\rho$ and Laplace transform $\chi$.
$I_0$ denotes the modified Bessel function of the first kind of order zero, 
$J_0$ is the Bessel function of the first kind of order zero, and $\delta(\cdot)$ is the Dirac delta function.}

\begin{tabular}{l|c|c|c|c}

\hline\hline
\multicolumn{1}{c|}{Type} & 
\multicolumn{1}{c|}{$\Phi(\varpi t)$} & 
\multicolumn{1}{c|}{$\rho(\theta)$} & 
\multicolumn{1}{c|}{$\chi(\Lambda,\varepsilon)$} & 
\multicolumn{1}{|c}{$\chi(i\Omega,\varepsilon)$}  \\

\hline\hline
&&&&\\
Sawtooth wave & 
$
\displaystyle{2\left(\frac{\varpi t}{2\pi}\, \mathrm{mod}\,\, 1\right)-1}
$ &
$
\left\{
\begin{array}{cc}
\frac{1}{2\varepsilon},& \theta\in[\tau_0-\varepsilon,\tau_0+\varepsilon]\\
0, &\text{elsewhere}
\end{array} \right.
$ &
$\displaystyle{\frac{\sinh(\Lambda\varepsilon)}{\Lambda\varepsilon}}$ &
$\displaystyle{\frac{\sin(\Omega\varepsilon)}{\Omega\varepsilon}}$ \\
&&&&\\
\hline

&&&&\\
Sine wave & 
$
\sin(\varpi t)
$ &
$\displaystyle{\frac{1}{\pi\sqrt{\varepsilon^2-(\theta-\tau_0)^2}}}$ &
$I_0(\Lambda\varepsilon)$ &
$J_0(\Omega\varepsilon)$ \\
&&&&\\

\hline
&&&&\\
Square wave & 
$
\mathrm{sgn}[\sin(\varpi t)]
$ &
$\displaystyle{\frac{\delta(\theta-\tau_0+\varepsilon)+\delta(\theta-\tau_0-\varepsilon)}{2}}$ &
$\cosh(\Lambda\varepsilon)$ &
$\cos(\Omega\varepsilon)$ \\
&&&&\\
\hline\hline
\end{tabular}
\end{table*}

With the ansatz $\delta\mathbf{\widetilde{x}}(t)=e^{\Lambda t} \mathbf{c}$, where $\mathbf{c}$ is a constant $d$-dimensional vector, we obtain a characteristic equation for $\Lambda=\Lambda(\nu)$  determining the stability of the collective fixed point:
\begin{align}
\mathrm{det}\left[\Lambda\widehat{\mathbf{I}}_d-\widehat{\mathbf{J}}
+\sigma_1\left(\mu-\nu\int_{0}^{\infty}\rho_1(\theta)e^{-\Lambda\theta}\,d\theta \right)\widehat{\mathbf{H}}_1\right.\nonumber\\
\left.
+\sigma_2\left(1-\int_{0}^{\infty}\rho_2(\theta)e^{-\Lambda\theta}\,d\theta \right)\widehat{\mathbf{H}}_2\right]=0.
\label{chareq1}
\end{align} 
The stability is clearly determined by the connection topology via the location of the eigenvalues $\nu$ of the interconnection adjacency matrix $\widehat{\mathbf{A}}$. Although the eigenspectrum of $\widehat{\mathbf{A}}$ depends on the concrete network, the boundaries of the spectrum can be succintly calculated by applying the Gershgorin's disc theorem \cite{EAR03,MIE12} that gives the region in the complex plane that contains all the eigenvalues of $\widehat{\mathbf{A}}$. Since $a_{jj}=0$ and $\sum_{n=1}^N a_{jn}=\mu$ for all $j$, all the eigenvalues of the adjacency matrix  $\widehat{\mathbf{A}}=\{a_{jn}\}$ lie within a disk of radius $\mu$ centered at the origin of the complex $\nu$-plane. Hence, the eigenvalues $\nu_j$ satisfy the condition:
\begin{equation}
|\nu_j|\leq\mu.
\end{equation}
If the interconnection adjacency matrix is symmetric, then the eigenvalues are real and, consequently, located in the interval $[-\mu,\mu]$. In this case, the maximum eigenvalue equals the row-sum of $\widehat{\mathbf{A}}$, i.e. $\nu_{max}=\mu$.

The stabilization of the network steady state $\mathbf{X}^*$ cannot always be achieved. From Eq. (\ref{chareq1}) we form a characteristic function
\begin{align}
H(\Lambda)=\mathrm{det}&\left[\Lambda\widehat{\mathbf{I}}_d-\widehat{\mathbf{J}}
+\sigma_1\left(\mu-\nu\int_{0}^{\infty}\rho_1(\theta)e^{-\Lambda\theta}\,d\theta \right)\widehat{\mathbf{H}}_1\right.\nonumber\\
&\left. +\sigma_2\left(1-\int_{0}^{\infty}\rho_2(\theta)e^{-\Lambda\theta}\,d\theta \right)\widehat{\mathbf{H}}_2\right].
\end{align}
Since one of the eigenvalues $\nu$ of the interconnection adjacency matrix $\widehat{\mathbf{A}}$ is equal to the row sum $\mu$ of $\widehat{\mathbf{A}}$, we focus on the case $\nu=\mu$. Considering the quasipolynomial form of the function $H(\Lambda)$ and restricting the dependence upon $\Lambda$ to the real axis, it is easy to show that  $H(\Lambda)>0$ in the limit $\Lambda\rightarrow\infty$. Also, in the case $\nu=\mu$, $H(0)=\det\left[-\widehat{\mathbf{J}}\right]=\prod_{j=1}^d (-s_j)$, where $s_j$ are the eigenvalues of the Jacobian matrix $\widehat{\mathbf{J}}$ of the local node dynamics. Consequently, if  $\widehat{\mathbf{J}}$ posesses an odd number of positive real eigenvalues, then $H(0)<0$, and there exists at least one positive real root of $H(\Lambda)=0$, meaning that the collective fixed point $\mathbf{X}^*$ is unstable for any values of the coupling parameters and any network topology. 

Another restriction for the fixed point stabilization occurs when the interaction is instantaneous and the self-feedback is absent, i.e. $\rho_{1,2}(\theta)=\delta(0)$, in which case the characteristic Eq. (\ref{chareq1}) is reduced to $\mathrm{det}[\Lambda\widehat{\mathbf{I}}_d-\widehat{\mathbf{J}}]=0$ for $\nu=\mu$. The stability in this case is completely determined by the eigenvalues of the Jacobian $\widehat{\mathbf{J}}$, meaning that unstable local dynamics induces instability in the connected network regardless of the connection topology and the coupling parameters. However, this restriction can be lifted by activating the self-feedback at the nodes (i.e. Pyragas-type control with a time-varying delay). 

In the present work, we consider a deterministic variation of the delay $\tau_1(t)$ in the interval $[\tau_{01}-\varepsilon_1,\tau_{01}+\varepsilon_1]$, and, correspondingly,  variation of $\tau_2(t)$ in the interval $[\tau_{02}-\varepsilon_2,\tau_{02}+\varepsilon_2]$. The specifications and the properties of the modulation types used in the analysis are provided in Table I.  In these cases, $\rho_{1,2}(\theta)$ are non-zero only in the interval of variation, and the characteristic equation for $\Lambda(\nu)$ simplifies to
\begin{align}
\mathrm{det}&\left[\Lambda\widehat{\mathbf{I}}_d-\widehat{\mathbf{J}}+\sigma_1\left(\mu-\nu e^{-\Lambda\tau_{01}}\chi_1(\Lambda,\varepsilon_1) \right)\widehat{\mathbf{H}}_1\right.\nonumber\\
&\left. +\sigma_2\left(1-e^{-\Lambda\tau_{02}}\chi_2(\Lambda,\varepsilon_2) \right)\widehat{\mathbf{H}}_2\right]=0,
\label{charmsf}
\end{align}
where 
\begin{align}
\chi_1(\Lambda,\varepsilon_1)&=\int_{-\varepsilon_1}^{\varepsilon_1}\rho_1(\tau_{01}+\theta)e^{-\Lambda\theta}\,d\theta,\nonumber\\
\chi_2(\Lambda,\varepsilon_2)&=\int_{-\varepsilon_2}^{\varepsilon_2}\rho_2(\tau_{02}+\theta)e^{-\Lambda\theta}\,d\theta
\end{align}
are the Laplace transforms of the associated distributed delay kernels (see Table I).

\section{The Stuart-Landau model}

We will analyze the conditions of amplitude death in an oscillator network whose dynamics on each node is described by a Stuart-Landau normal form equation. The Stuart-Landau system describes a generic limit cycle oscillator that shares a common local dynamics with real oscillator systems where periodic state arise from a fixed point through a Hopf bifurcation. 
We consider a network of delay-coupled Stuart-Landau oscillators with time-varying delayed coupling and self-feedback:
\begin{align}
\dot z_j=h(z_j)+\sigma_1e^{i\beta_1}\sum_{n=1}^{N}a_{jn}\left[z_n(t-\tau_1(t))-z_j(t)\right]\nonumber\\
+\sigma_2e^{i\beta_2}\left[z_j(t-\tau_2(t))-z_j(t)\right]
\label{sys1}
\end{align}
with $z_j\in\mathbb{C}$ and $j=1,2\dots N$. 
The local dynamics is given by the normal form 
\begin{equation}
h(z_j)=\left[\lambda+i\omega\mp(1+i\gamma)|z_j|^2\right]z_j,
\label{sys2}
\end{equation}
where the minus (plus) sign corresponds to a supercritical (subcritical) Hopf bifurcation. The parameters $\sigma_{1,2}$ are the coupling strengths, and $\beta_{1,2}$ are the coupling phases.  
The state variable $z_j$ of an individual oscillator is given by $z_j=r_j e^{i\varphi_j}$ in polar coordinates, and by $z_j=x_j+i y_j$ in rectangular coordinates.
In the absence of any interaction ($\sigma_{1,2}=0$), the dynamics of each individual oscillator in polar representation is given by $\dot r_j=(\lambda\mp r_j^2)r_j$ and $\dot\varphi_j=\omega\mp\gamma r_j^2$.
The uncoupled system has a fixed point at $r_j=0$, which is stable if $\lambda<0$ and unstable if $\lambda>0$ in both the supercritical and subcritical regime. The Hopf bifurcation occurs at the critical point $\lambda=0$. The system also has a stable periodic orbit $r_j=\sqrt{\lambda}$ in the supercritical case for $\lambda>0$ with a period $T=2\pi/(\omega-\gamma\lambda)$, and an unstable periodic orbit $r_j=\sqrt{-\lambda}$ in the subcritical case  for $\lambda<0$ with the same period $T$.

In the following, we focus on the stability of the fixed point $r_j=0$ of the coupled dynamics, i.e., the regime of amplitude death. We assume $\lambda>0$, i.e., the origin is unstable in the uncoupled system, and investigate the influence of the variable-delay coupling and the network topology on its stability. We consider the limiting case of high-frequency modulations of the delays $\tau_{1,2}(t)$, and use the distributed delay limit analysis in determining the stability of the steady state. The system (\ref{sys1}) in rectangular coordinates takes the form of Eq. (\ref{system}), where $\mathbf{x}_j(t)=(x_j(t),y_j(t))^T$ is the two-dimensional state vector ($d=2$), 
\begin{equation}
\mathbf{f}(\mathbf{x})=
\left(
\begin{array}{c}
x\left[\lambda\mp(x^2+y^2)\right]-y\left[\omega\mp\gamma(x^2+y^2)\right]\\
x\left[\omega\mp\gamma(x^2+y^2)\right]+y\left[\lambda\mp(x^2+y^2)\right]\\
\end{array}\right)
\end{equation}
is the vector field of the node dynamics, and
\begin{equation}
\widehat{\mathbf{H}}_{1,2}=
\left(
\begin{array}{cc}
\cos\beta_{1,2} & -\sin\beta_{1,2} \\
\sin\beta_{1,2} & \cos\beta_{1,2} \\
\end{array}
\right)
\label{rotmatrix}
\end{equation}
are $2\times 2$ rotational matrices related to the phase-dependent coupling in the interconnection $(\widehat{\mathbf{H}}_1)$ and in the self-feedback ($\widehat{\mathbf{H}}_2)$. The collective fixed point of the network is a $2N$-dimensional null vector whose stability is determined by the characteristic Eq. (\ref{charmsf}) with the corresponding Jacobian matrix 
\begin{equation}
\widehat{\mathbf{J}}=
\left(
\begin{array}{cc}
\lambda & -\omega \\
\omega & \lambda \\
\end{array}
\right).
\label{jacobian}
\end{equation}
The resulting equation for $\Lambda(\nu)$ is simplified into
\begin{align}
\Lambda+\sigma_1e^{\pm i\beta_1}\left[\mu-\nu e^{-\Lambda\tau_{01}}\chi_1(\Lambda,\varepsilon_1)\right]&\nonumber\\
+\sigma_2e^{\pm i\beta_2}\left[1- e^{-\Lambda\tau_{02}}\chi_2(\Lambda,\varepsilon_2)\right]&=\lambda\pm i\omega.
\label{charmsfsl}
\end{align}
To obtain the parametric representation of the boundaries of stability of the master stability function in the complex $\nu$-plane, we substitute $\Lambda=i\Omega$ and $\nu=p+iq$ ($p=\mathrm{Re}(\nu), q=\mathrm{Im}(\nu)$) into Eq. (\ref{charmsfsl}) and separate the resulting complex equation into two real-valued equations:
\begin{align}\label{eq:bound_stab}
\sigma_1\left[\mu\cos\beta_1-\chi_1(p\cos\psi_1+q\sin\psi_1)\right]&\nonumber\\
+\sigma_2\left(\cos\beta_2-\chi_2\cos\psi_2\right)&=\lambda,\\
\Omega+\sigma_1\left[\pm\mu\sin\beta_1+\chi_1(p\sin\psi_1-q\cos\psi_1)\right]&\nonumber\\
+\sigma_2\left(\pm\sin\beta_2+\chi_2\sin\psi_2\right)&=\pm\omega,
\end{align} 
where for compactness we use the abbreviations
\begin{align}
&\chi_1=\chi_1(i\Omega,\varepsilon_1),  &\chi_2=\chi_2(i\Omega,\varepsilon_2),\\
&\psi_1=\Omega\tau_{01}\mp\beta_1, &\psi_2=\Omega\tau_{02}\mp\beta_2.
\end{align}
We have taken into account that for the delay variations used in our simulations, the distributed-delay kernels $\rho_{1,2}(\theta)$ are even functions around the corresponding mean delays, meaning that the associated functions $\chi_{1,2}$ are real-valued functions (see Table I).
By algebraically manipulating the system Eqs.(\ref{eq:bound_stab}), we obtain the parametric dependence of the stability crossing curves in the complex $(p,q)$ plane on the intrinsic eigenfrequency $\Omega$:
\begin{align}
p&(\Omega)_{1,2}=\frac{1}{\sigma_1\chi_1}\times\nonumber\\
\times&\left\{\left[\sigma_1\mu\cos\beta_1-\lambda+\sigma_2(\cos\beta_2-\chi_2\cos\psi_2)\right]\cos\psi_1\right.\nonumber\\
+&\left.\left[\pm\omega-\Omega\mp \sigma_1\mu\sin\beta_1+\sigma_2(\mp\sin\beta_2-\chi_2\sin\psi_2)\right]\sin\psi_1\right\},
\label{bound1}
\nonumber\\ \\
q&(\Omega)_{1,2}=\frac{1}{\sigma_1\chi_1}\times\nonumber\\
\times&\left\{\left[\sigma_1\mu\cos\beta_1-\lambda+\sigma_2(\cos\beta_2-\chi_2\cos\psi_2)\right]\sin\psi_1\right.\nonumber\\
-&\left.\left[\pm\omega-\Omega\mp \sigma_1\mu\sin\beta_1+\sigma_2(\mp\sin\beta_2-\chi_2\sin\psi_2)\right]\cos\psi_1\right\}.
\nonumber\\
\label{bound2}
\end{align}
A subset of the stability crossing curves given by Eqs. (\ref{bound1})-(\ref{bound2}) describe the stability boundary of the master stability function in the complex $\nu$-plane.

\subsection{Case I: No self-feedback ($\sigma_2=0$)}

Throughout the rest of the paper, we take the parameters of the uncoupled system as $\lambda=0.1$, $\omega=1$ and $\gamma=0.1$, for which the steady state at the origin is unstable.
We first analyze the stability of the steady state in the absence of self-feedback ($\sigma_2=0$). 
We re-write the characteristic Eq. (\ref{charmsfsl}) as:
\begin{equation}
\Lambda+\sigma_1e^{\pm i\beta_1}\left[\mu-|\nu| e^{-i\varphi} e^{-\Lambda\tau_{01}}\chi_1(\Lambda,\varepsilon_1)\right]=\lambda\pm i\omega,
\label{slcharnoself}
\end{equation}
where we employed the complex representation $\nu=|\nu|e^{-i\varphi}$ for the eigenvalues of the interconnection adjacency matrix. The stability crossing curves are obtained when $\Lambda=i\Omega$, and we have:
\begin{align}
\sigma_1\left[\mu\cos\beta_1-|\nu|\chi_1\cos(\Omega\tau_{01}\mp\beta_1-\varphi)\right]&=\lambda,
\label{stabcq1}\\
\sigma_1\left[\pm\mu\sin\beta_1+|\nu|\chi_1\sin(\Omega\tau_{01}\mp\beta_1-\varphi)\right]&=\pm\omega-\Omega,
\label{stabcq2}
\end{align}
from which we obtain:
\begin{equation}
|\nu|^2=\frac{(\sigma_1\mu\cos\beta_1-\lambda)^2+(\pm\omega-\Omega\mp \sigma_1\mu\sin\beta_1)^2}{{\sigma_1}^2{\chi_1}^2}.
\label{dist}
\end{equation}
Taking into account that $|\chi_1|\leq1$, we can make a crude estimate of the minimum radius $|\nu|_{\mathrm{min}}$ of the circular region centered at the origin in the complex $\nu$-plane that does not contain any stability crossing curves: 
\begin{equation}
|\nu|_{\mathrm{min}}=\left|\mu\cos\beta_1-\frac{\lambda}{\sigma_1}\right|.
\label{disk}
\end{equation}
From Eq. (\ref{dist}) we see that the distance of the stability crossing curves from the origin increases as soon as $\varepsilon_1>0$ due to the fact that $|\chi_1|\leq1$. Consequently, the disk region $|\nu|\leq|\nu|_{\mathrm{min}}$ is generally enlarged by introducing variable delays. Depending on whether this disk encloses stability or instability region, the enlargement could have a positive or negative effect on the stability of the collective fixed point.
To elucidate the result, we consider the point $|\nu|$=0, i.e., the origin of the complex $\nu$-plane. From Eq. (\ref{slcharnoself}) we see that this point is characterized by a complex-conjugate pair of characteristic exponents, whose real and imaginary parts are given by:
\begin{align}
\mathrm{Re}(\Lambda)&=\lambda-\sigma_1\mu\cos\beta_1,\label{relambda}\\
\mathrm{Im}(\Lambda)&=\pm(\omega-\sigma_1\mu\sin\beta_1).
\end{align}
The shape of the (in)stability region can be deduced by considering the change of $\mathrm{Re}(\Lambda)$ at the stability crossing curves $\Lambda=i\Omega$. From Eq. (\ref{slcharnoself}), in the constant delay case $\chi_1=1$, we obtain:
\begin{equation}
\mathrm{sign\,Re}\left(\frac{d\Lambda}{d|\nu|}\right)_{\Lambda=i\Omega}=\sigma_1^2\tau_{01}|\nu|^2+\sigma_1\mu\cos\beta_1-\lambda.
\label{sign}
\end{equation}
We first consider the case $\mathrm{Re}(\Lambda)<0$, corresponding to the choice of the system parameter values that fulfill the condition
\begin{equation}
\lambda-\sigma_1\mu\cos\beta_1<0.
\label{stabilitycondition}
\end{equation}
Since the stability crossing curves in this case are not contained within the disk (\ref{disk}), each point within this disk possesses an infinite number of characteristic exponents with negative real parts, i.e., the disk Eq.~(\ref{disk}) encloses a stability region. In this case, the sign in Eq.~(\ref{sign}) is always positive, which means that at each stability crossing curve outside this disk, the characteristic exponents $\Lambda$ are crossing the imaginary axis moving from the left to the right in the complex $\Lambda$-plane. Consequently, in this case the stability region is a connected set in the complex $\nu$-plane containing the disk Eq.~(\ref{disk}). If all the eigenvalues $\nu$ of the adjacency matrix are located within this region, the collective steady state will be stable, and we have achieved amplitude death. On the other hand, if $\lambda-\sigma_1\mu\cos\beta_1>0$, which is the case $\mathrm{Re}(\Lambda)>0$ in Eq.~(\ref{relambda}), each point within the disk Eq.~(\ref{disk}) possesses an 
infinite set of characteristic exponents with positive real parts, i.e. the disk encloses a part of an instability region. While the delay modulation in the previous case enlarges the stability region of the master stability function, in this case modulation enlarges the instability region instead. 

\begin{figure}
\includegraphics[width=\columnwidth,height=!]{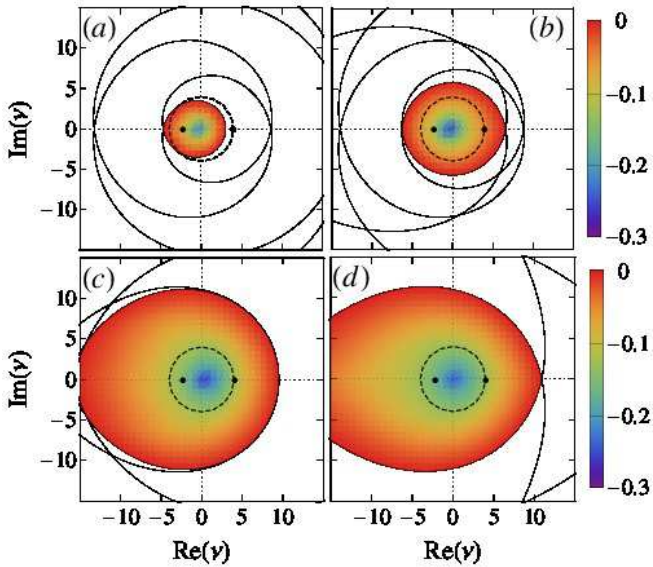}
\caption{(Color online) Master stability function in the complex
$\nu$ plane for a $2k$-ring network of Stuart-Landau oscillators without self-feedback. Each oscillator is bidirectionally coupled to four of its nearest neighbors with $k=2$ connections on each side, i.e., the network has a constant node degree $\mu=2k=4$. The stability region is calculated from Eq. (\ref{slcharnoself}) for a sawtooth-wave modulation of the delay in the high-frequency regime (i.e., distributed delay limit) around a mean value $\tau_{01}=2\pi$ for different modulation amplitudes $\varepsilon_1$:  (a) $\varepsilon_1=0$ (constant delay), (b) $\varepsilon_1=2$, (c) $\varepsilon_1=4$, (d) $\varepsilon_1=6$. The other parameters are: $\lambda=0.1$, $\omega=1$, $\gamma=0.1$, $\beta_1=0$, $\sigma_1=0.1$. The stability region is color coded by the maximum negative real part of the eigenvalues $\Lambda$. The stability region is bounded by a subset of the stability crossing curves Eqs.~(\ref{bound1})--(\ref{bound2}) (solid lines). The Gershgorin circle $|\nu|=\mu=4$ is dashed, and the solid black dots are the maximum and minimum eigenvalues of the network adjacency matrix for $N=20$ oscillators:  $\nu_{\mathrm{min}}\approx-2.236$, $\nu_{\mathrm{max}}=4$.}
\label{fig1}
\end{figure}
\begin{figure}
\includegraphics[width=\columnwidth,height=!]{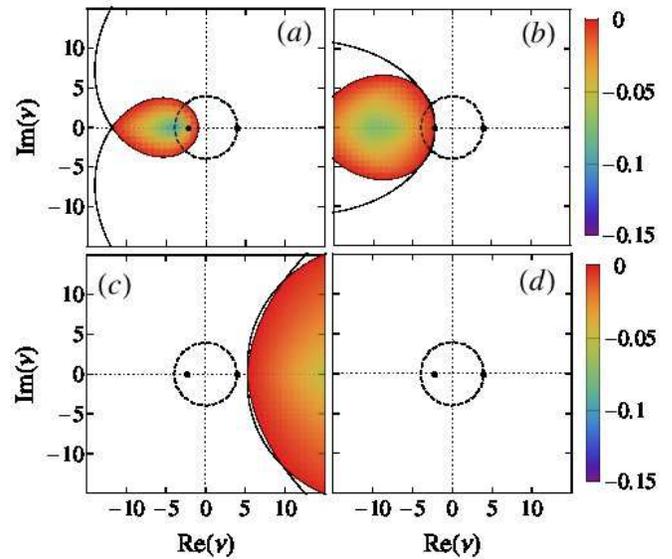}
\caption{(Color online) Master stability function in the complex $\nu$ plane corresponding to Fig. \ref{fig1}, with $\sigma_1=0.02$ and the other parameters unchanged.}
\label{fig2}
\end{figure}

To confirm the results, we have numerically analyzed the characteristic Eq. (\ref{slcharnoself}). Figure \ref{fig1} shows the master stability function of the Stuart-Landau oscillator network without self-feedback. The topology of the network is a ring with a constant node degree $\mu=4$. Each oscillator is bidirectionally coupled to $\mu=2k=4$ nearest neighbors with $k=2$ connections (edges) on each side.
The delay time $\tau_1(t)$ in the interconnection is modulated with a sawtooth-wave around a mean delay value $\tau_{01}=2\pi$.
The other coupling parameters are set to $\beta_1=0$, and $\sigma_1=0.1$. The parameter values at which the master stability function is negative is denoted by the color-shaded (grey-shaded) region, and the color code corresponds to the largest negative real part of the characteristic exponents $\Lambda$ calculated from Eq.~(\ref{slcharnoself}).  Different panels correspond to different values of the modulation amplitude $\varepsilon_1$.
The Gershgorin circle $|\nu|=\mu=4$ is denoted by a dashed line, and the solid black dots are the maximum and minimum eigenvalues of the adjacency matrix for a $2k$-ring network with $N=20$ oscillators calculated from Eq.~(\ref{eigenvalues}) in the Appendix. The adjacency matrix of the ring topology is symmetric (see Appendix), and thus has a real eigenspectrum contained in the Gershgorin interval $[-4,4]$. In the constant delay case in panel (a), the maximum eigenvalue $\nu_{\mathrm{max}}=4$ lies outside the stability domain, rendering the collective fixed point unstable. As the modulation amplitude increases (panels (b)--(d)), the stability region is monotonically enlarged, and eventually surpasses the Gershgorin disk, thus stabilizing the fixed point. The resulting enlargement of the stability region is in agreement with the analysis in the previous paragraph, since in this case the condition (\ref{stabilitycondition}) is fulfilled. 
Consequently, the criterion (\ref{stabilitycondition}) is a necessary condition for amplitude death for the considered $2k$-ring network topology.
The stability region in the opposite situation, when $\lambda-\sigma_1\mu\cos\beta_1>0$, is depicted in Fig. \ref{fig2} for $\sigma_1=0.02$ and the rest of the parameters unchanged. Accordingly, it is seen that delay modulation enlarges the instability region in this case, and achieving amplitude death becomes impossible for any modulation amplitude as confirmed by our previous analysis.

\begin{figure*}
\includegraphics[width=0.8\textwidth,height=!]{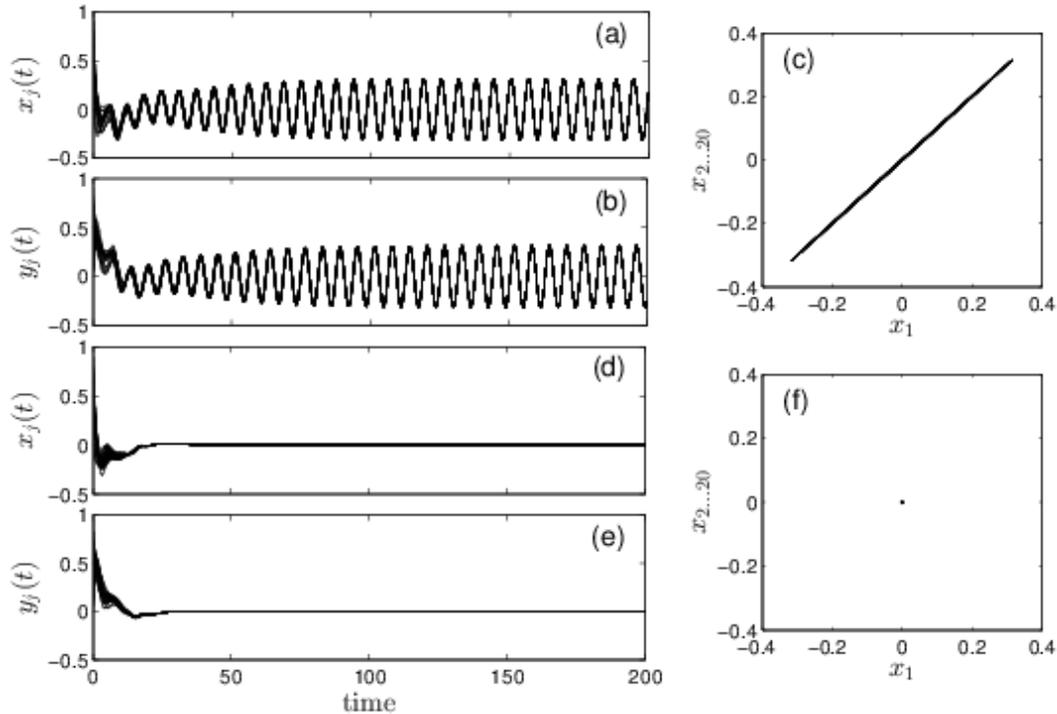}
\caption{Controlling amplitude death by variable-delay interconnections in a system of $N=20$ Stuart-Landau oscillators in a regular ring network with $k=2$ links on each side of a node ($\mu=4$) without self-feedback. 
(a),(b): Time series of the variables $x_j=\{x_1,x_2\dots x_{20}\}$ and $y_j=\{y_1,y_2\dots y_{20}\}$ for constant delay, and the associated synchronization diagram $x_1$ vs. $x_{2,\dots, 20}$ [panel (c)] indicating in-phase synchronous periodic dynamics. 
(d),(e): Time series of  $x_j$ and $y_j$ for a sawtooth-wave delay modulation with amplitude $\varepsilon_1=2\pi$ and frequency $\varpi=10$, stabilizing the unstable origin, i.e., giving rise to amplitude death [panel (f)]. 
Parameters: $\lambda=0.1$, $\omega=1$, $\gamma=0.1$, $\beta_1=0$, $\tau_{01}=2\pi$, $\sigma_1=0.1$ and $\sigma_2=0$.
The simulations were performed using the MATLAB routine \texttt{ddesd} for integrating delay-differential equations with general delays.}
\label{simulation}
\end{figure*}

To verify the successful stabilization of the origin by including variable delays, we have performed computer simulations of the network dynamics by numerically integrating the system Eqs.~(\ref{sys1})--(\ref{sys2}). The resulting diagrams are shown in Fig. \ref{simulation}. The coupling parameters in each panel are chosen as $\beta_1=0$, $\tau_{01}=2\pi$, $\sigma_1=0.1$ and $\sigma_2=0$. At these parameter values, the fixed point stabilization is unsuccessful in the constant delay case, but amplitude death can be achieved if the delay is modulated. This result has already been implied from the analysis of the stability regions depicted in Fig. \ref{fig1}, and our simulations confirm this analysis. Panels (a) and (b) depict the time series of the system variables $x_j$ and $y_j$ ($j=1\dots 20$), respectively, in a constant delay case. The simulations are performed in a time-window of 200 time units for $N=20$ Stuart-Landau oscillators in a regular $2k$-ring network with $k=2$. It is observed that after a 
short transient, the system becomes synchronized without a phase lag (in-phase synchronization), which is confirmed by the synchronization diagram depicted in panel (c). Applying a modulated delay in form of a sawtooth wave with $\varepsilon_1=2\pi$ and $\varpi_1=10$ results in amplitude death, as is seen from the time series in panels (d) and (e) and the corresponding synchronization diagram in panel (f).  

\begin{figure*}
\hspace{-0.5cm}(a)\includegraphics[width=0.55\columnwidth,height=!]{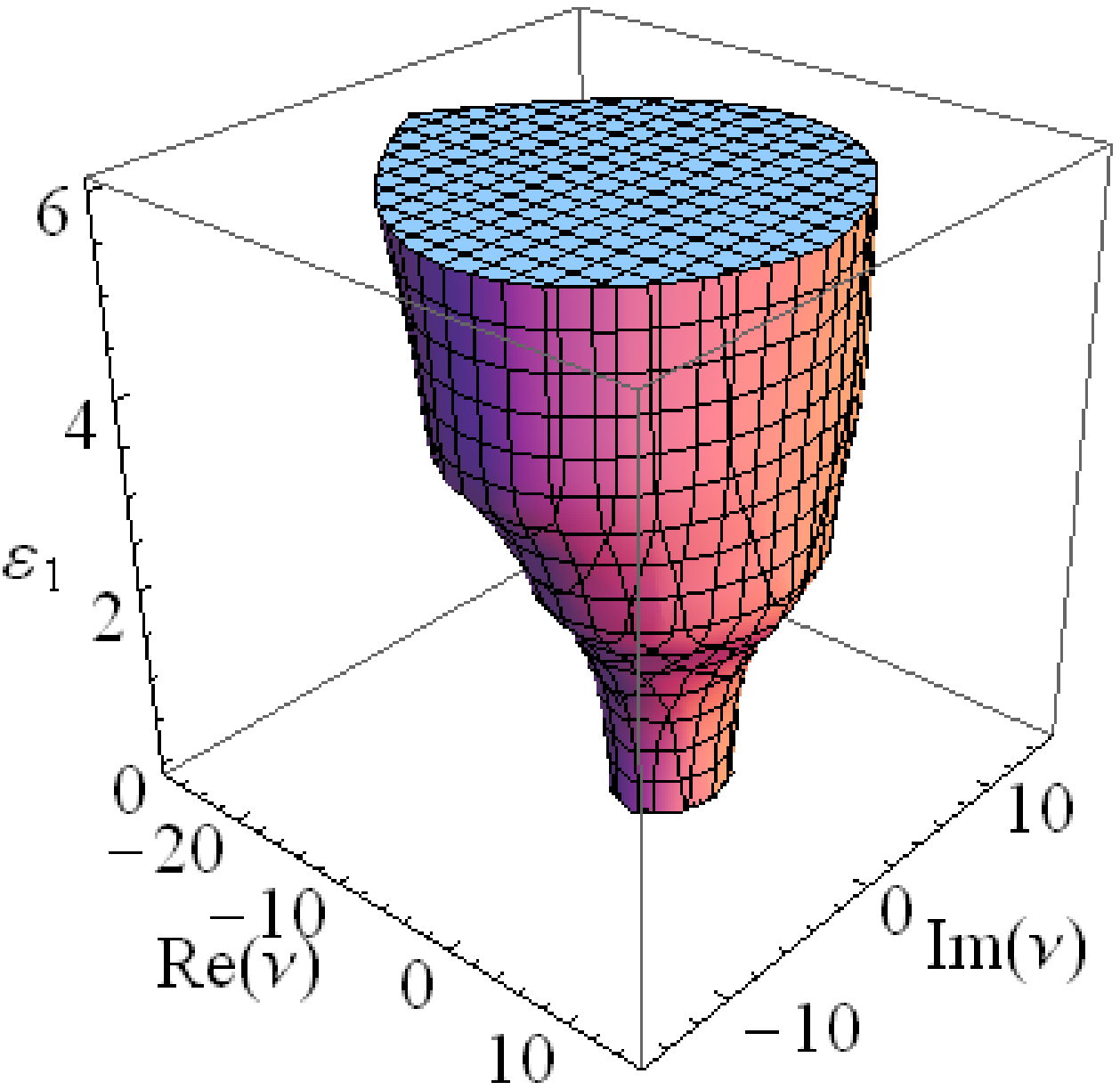}\hspace{0.5cm}
(b)\includegraphics[width=0.55\columnwidth,height=!]{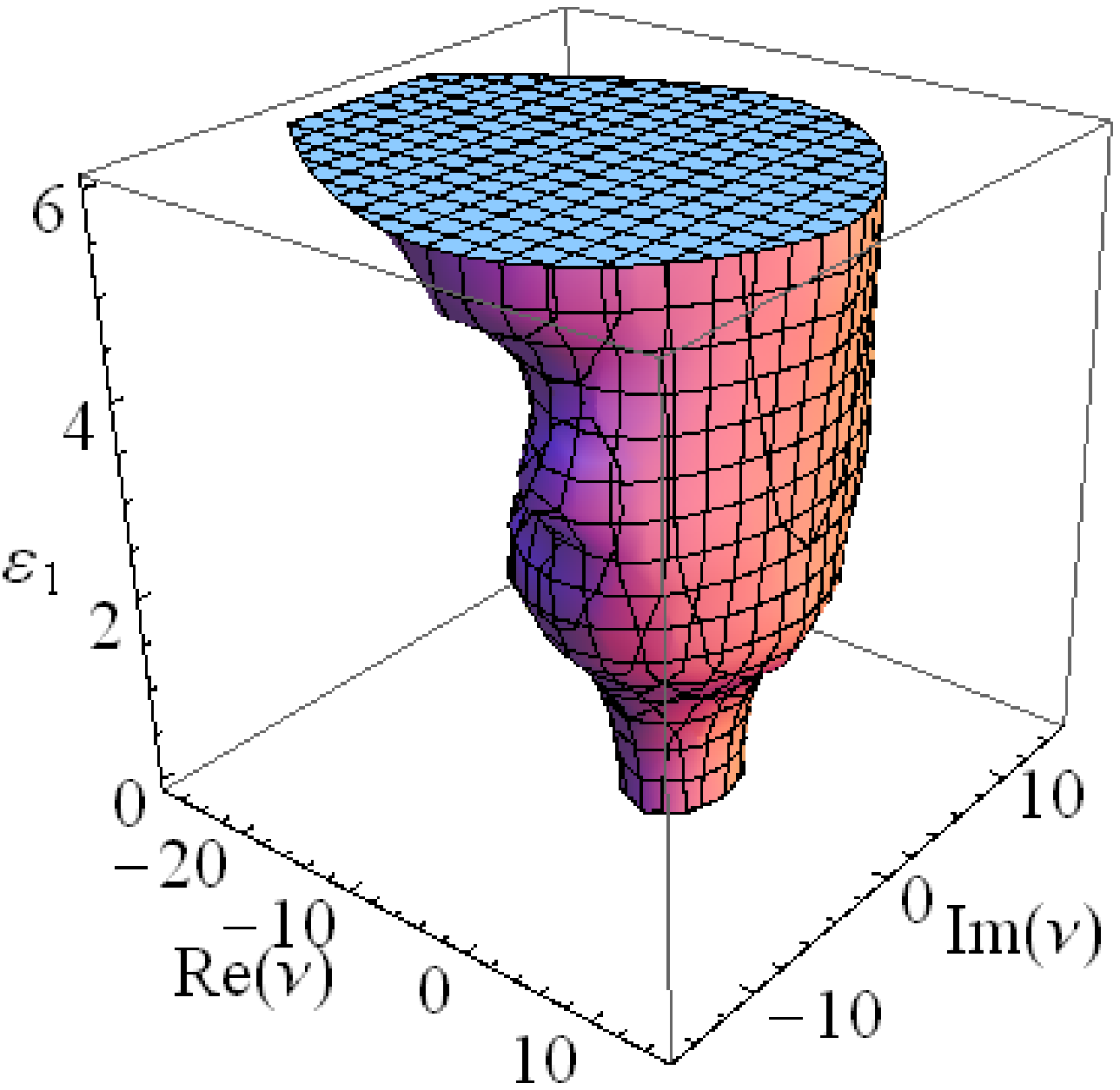}\hspace{0.5cm}
(c)\includegraphics[width=0.55\columnwidth,height=!]{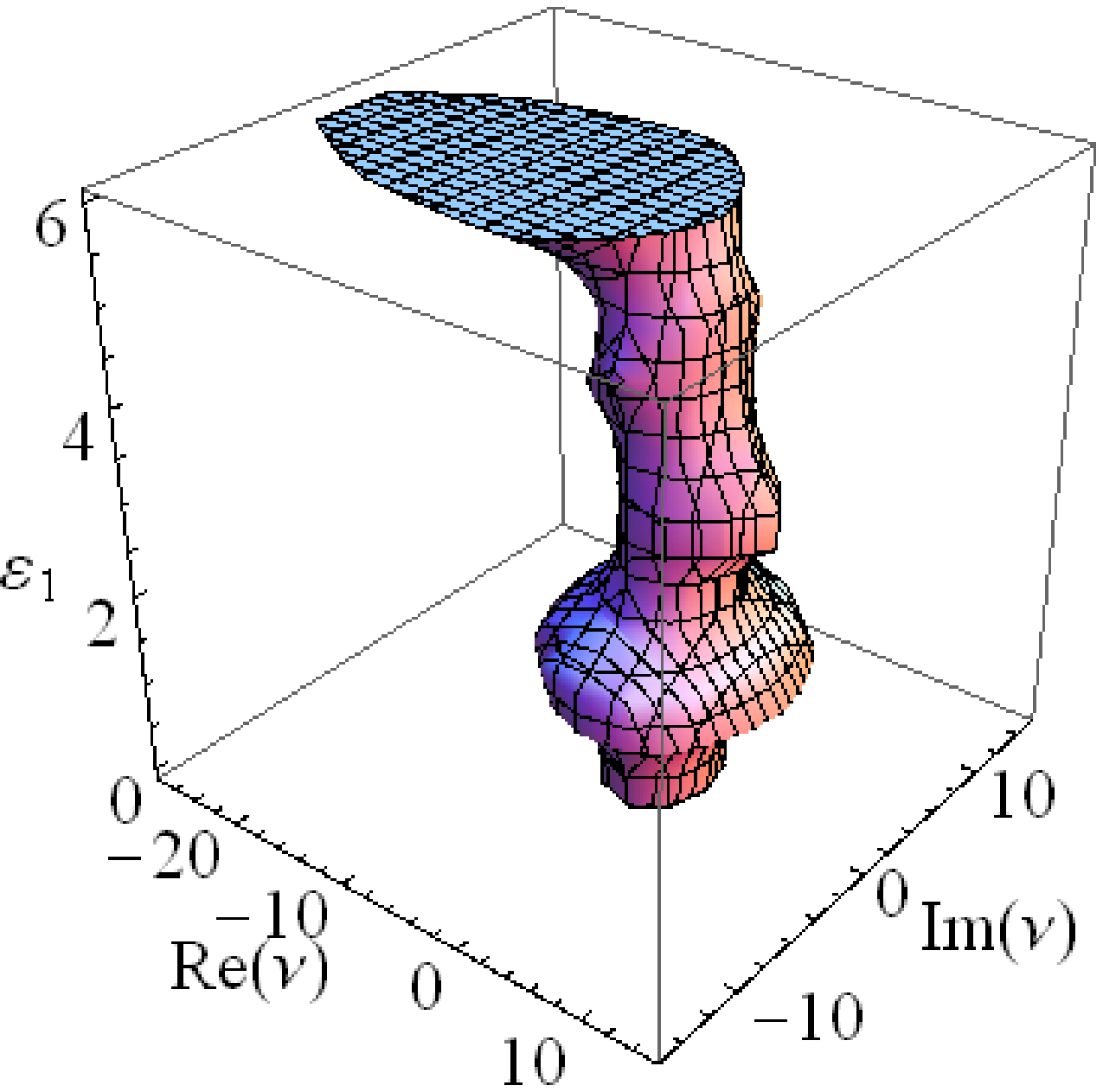}\\
(d)\includegraphics[width=0.6\columnwidth,height=!]{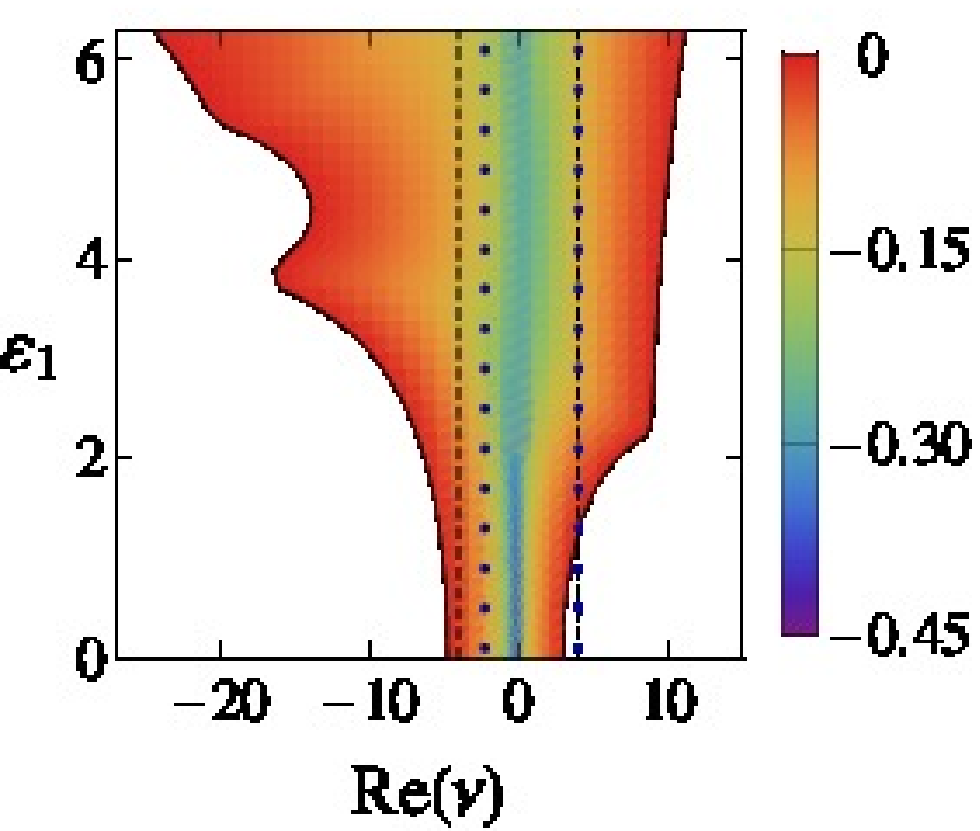}\hspace{0.15cm}
(e)\includegraphics[width=0.6\columnwidth,height=!]{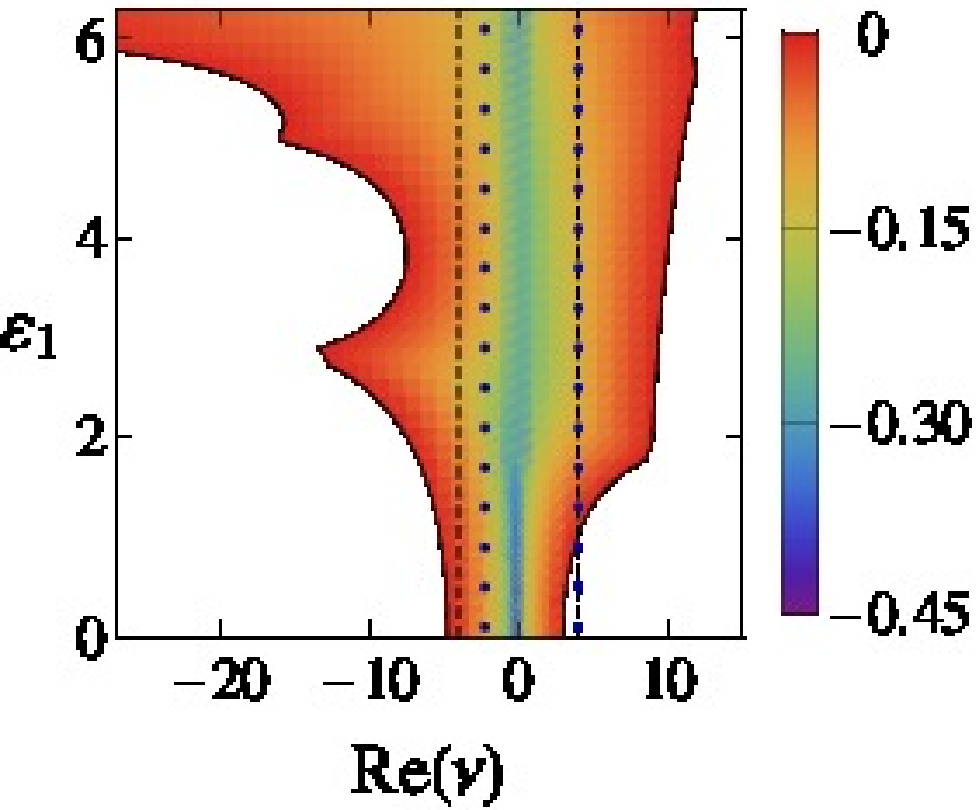}\hspace{0.15cm}
(f)\includegraphics[width=0.6\columnwidth,height=!]{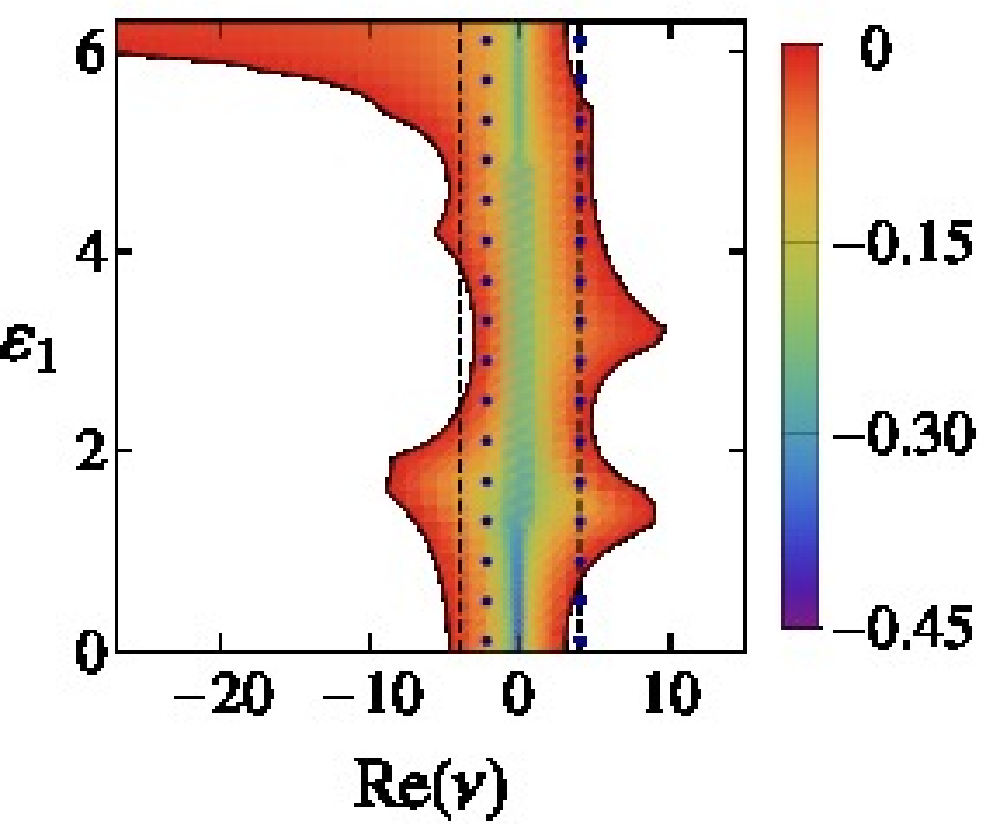}
\caption{(Color online) Master stability function without self-feedback in dependence of the modulation amplitude $\varepsilon_1$ for different types of delay modulations: (a, d) sawtooth-wave modulation; (b, e) sine-wave modulation; (c, f) square-wave modulation.  Panels (d)--(f) are sections of (a)--(c) at $\mathrm{Im}(\Lambda)=0$. The Gershgorin interval $|\mathrm{Re}(\nu)|\leq\mu= 2k$ is contained between the black dashed lines, and the blue dotted lines correspond to the maximum and minimum eigenvalues of the adjacency matrix for $2k$-ring network with $N=20$ oscillators and $k=2$ connections on each node side. Other parameters are as in Fig. \ref{fig1}.}
\label{fig3}
\end{figure*}

Achieving amplitude death by enlarging the stability region of the master stability function via variable delay interconnections is also observed for other types of delay modulation as long as the stability criterion (\ref{stabilitycondition}) is fulfilled. For comparison purposes, in Fig. \ref{fig3} we show the stability regions of the master stability function in dependence on the modulation amplitude $\varepsilon_1$ for sawtooth-wave (panels a, d), sine-wave (panels b, e) and square-wave modulation of the delay (panels c, f). The system parameters are the same as in Fig. \ref{fig1}. Panels (a)--(c) are the three-dimensional plots of the stability region, and panels (d)--(f) are vertical cuts of the corresponding three-dimensional representations at $\mathrm{Im}(\Lambda)=0$. The latter two-dimensional plots are relevant for the $2k$-ring network topology considered in this paper, since in this case the eigenvalues $\nu$ are real. In panels (d)--(f) the Gershgorin interval is bounded by the black dashed 
lines, and the blue dotted lines correspond to the maximum and minimum eigenvalues of the network adjacency matrix for $N=20$ oscillators ($\nu_{\mathrm{min}}\approx-2.236$, $\nu_{\mathrm{max}}=4$). Note that applying a modulated delay results in amplitude death for specific ranges of the modulation amplitude $\varepsilon_1$, when all the eigenvalues of the adjacency matrix are located inside the stability region. The enlargement of the stability region is more pronounced and almost monotonic for sawtooth-wave and sine-wave modulations, whereas this behavior is rather non-monotonic for a square-wave modulation, in which case the $\varepsilon_1$-intervals at which amplitude death occurs may even be disconnected. 

\begin{figure}
\includegraphics[width=\columnwidth,height=!]{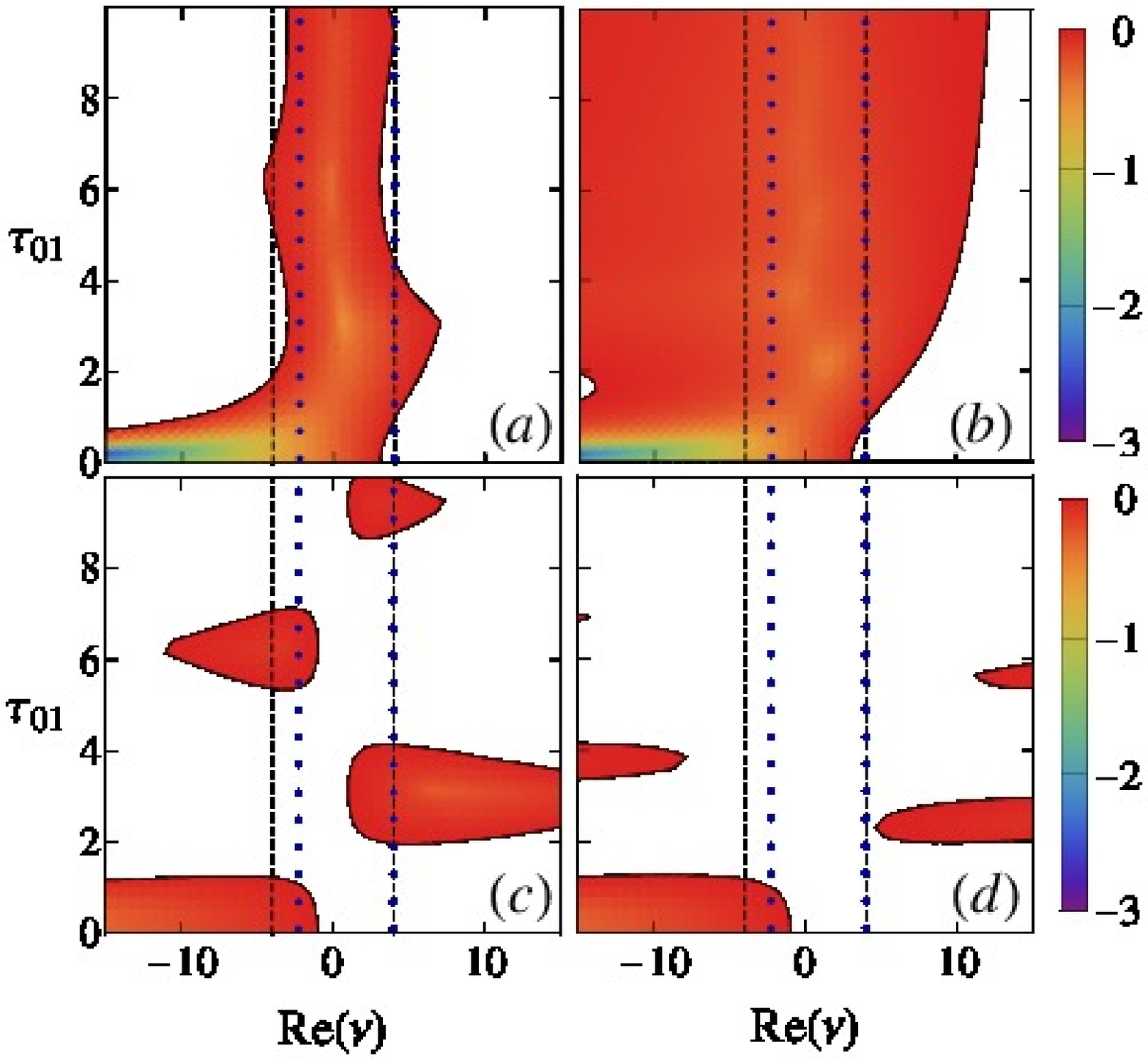}
\caption{(Color online) Master stability function in the $(\tau_{01},\mathrm{Re}(\nu))$ plane. (a) $\sigma_1=0.1$, $\varepsilon_1=0$ (no modulation). (b) $\sigma_1=0.1$, $\varepsilon_1=\tau_{01}$ (sawtooth-wave modulation). (c) $\sigma_1=0.02$, $\varepsilon_1=0$ (no modulation). (b) $\sigma_1=0.02$, $\varepsilon_1=\tau_{01}$ (sawtooth-wave modulation). The rest of parameters as in Fig. \ref{fig1}.}
\label{fig4}
\end{figure}

To investigate the dependence of the stability region on the mean delay $\tau_{01}$, in Fig. \ref{fig4} we show the master stability function in the  $(\tau_{01},\mathrm{Re}(\nu))$ plane for a sawtooth-wave modulation at different modulation amplitudes. As before, we consider a ring network with $N=20$ and $k=2$. The Gershgorin interval is marked by the black dashed lines, and the eigenspectrum of the adjacency matrix is contained within the interval bounded by the blue dotted lines. Panel (a) depicts the stability region for $\sigma_1=0.1$ in the constant delay case ($\varepsilon_1=0$), for which the stability criterion in Eq. (\ref{stabilitycondition}) is fulfilled. It is evident that the intervals for $\tau_{01}$ that warrant amplitude death are narrow and disconnected. By including variable delays, the stability region expands significantly, and the amplitude death now occurs in wide $\tau_{01}$ interval. The case $\varepsilon_1=\tau_{01}$ is shown in panel (b). In panels (c) and (d) we depict the 
corresponding master stability function for $\sigma_1=0.02$ for which Eq. (\ref{stabilitycondition}) is not satisfied. In this case, it can be seen that amplitude death cannot be achieved in the constant delay case for any value of the delay time (panel (c)), and by introducing variable delays and enlarging the modulation amplitude the instability region becomes even more expanded (panel (d)).

\begin{figure}
\includegraphics[width=\columnwidth,height=!]{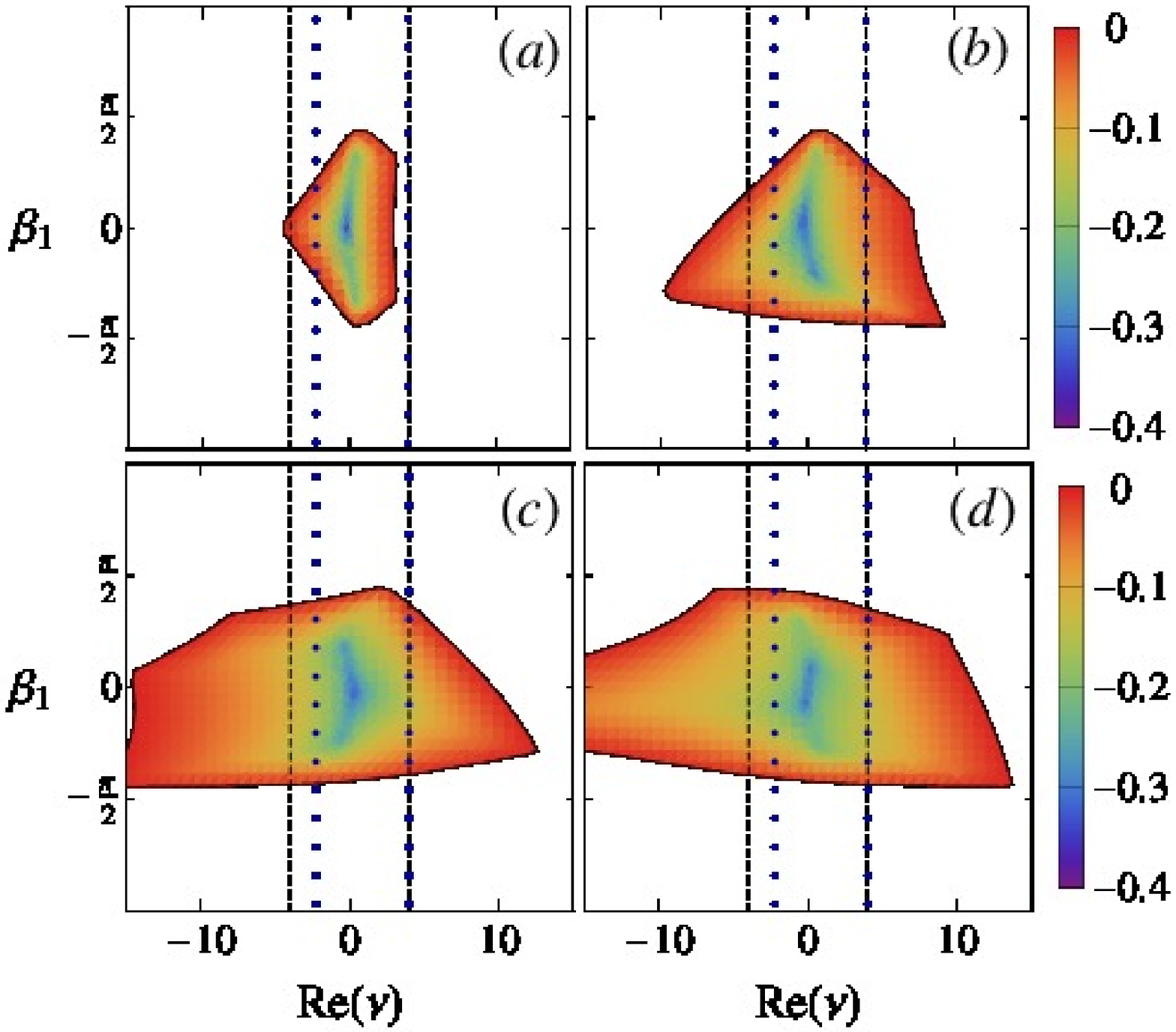}
\caption{(Color online) Master stability function in the $(\beta_1,\mathrm{Re}(\nu))$ plane for a sawtooth-wave delay modulation and increasing modulation amplitude: (a) $\varepsilon_1=0$, (b) $\varepsilon_1=2\pi/3$, (c) $\varepsilon_1=4\pi/3$, (d) $\varepsilon_1=2\pi$. Other parameters as in Fig. \ref{fig1}.}
\label{fig5}
\end{figure}

It is also desirable to investigate the influence of the variable delays on the stability region in the plane spanned by the coupling phase $\beta_1$ and $\mathrm{Re}(\nu)$. In Fig. \ref{fig5} we show the corresponding master stability function for a sawtooth-wave modulation of the delay $\tau_{01}$ at different modulation amplitudes $\varepsilon_1$.
The coupling parameters are $\tau_{01}=2\pi$ and $\sigma_1=0.1$, and again we consider a $2k$-ring network with $N=20$ and $k=2$. Panel (a) depicts the situation without delay modulation, i.e., $\varepsilon_1=0$.  In this case, the maximum eigenvalue $\nu_{\mathrm{max}}=4$ lies outside the stability region for any $\beta_1$, and amplitude death cannot be achieved. As the modulation amplitude increases (panels (b)--(d)), the stability region expands correspondingly, and amplitude death becomes possible at a fixed connected interval for $\beta_1$ depending on the modulation amplitude. 

\begin{figure}
\includegraphics[width=\columnwidth,height=!]{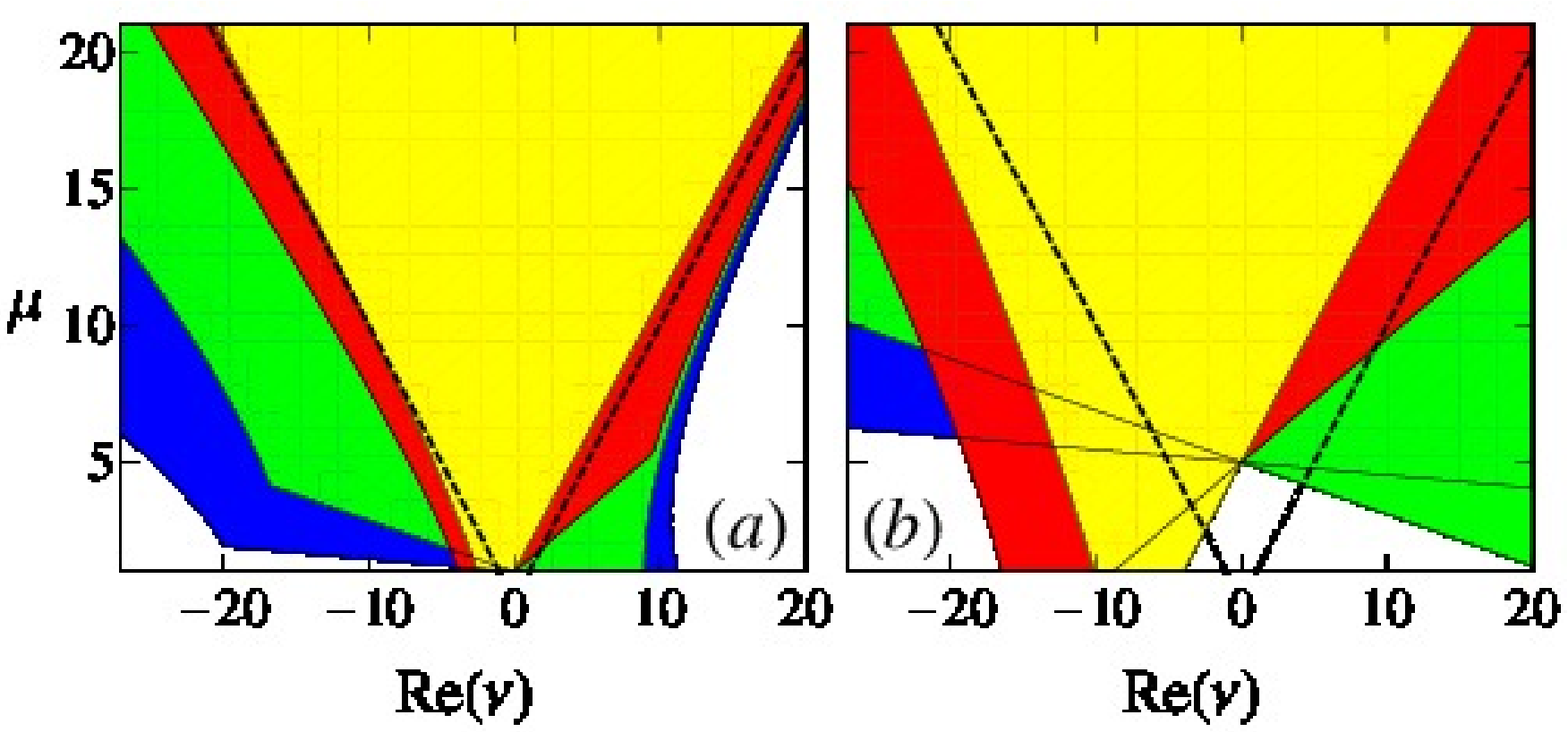}
\caption{(Color online) Master stability function in the  $(\mu,\mathrm{Re}(\nu))$ plane for a sawtooth-wave modulation of the delay around the mean value $\tau_{01}=2\pi$. The yellow, red, green, and blue (light gray, dark gray, medium gray, and black) domains correspond to $\varepsilon_1=0$, $2$, $4$, and $6$, respectively. Gain parameter: (a) $\sigma_1=0.1$, (b) $\sigma_2=0.02$. Other parameters as in Fig. \ref{fig1}.}
\label{muchanged}
\end{figure}

To see that the enlargement of the region of the master stability function is not restricted only to the case $\mu=4$ considered above, we have calculated the stability region depending on the node degree $\mu$ in case of a real eigenspectrum of the interconnection adjacency matrix, e.g., the previously considered bidirectional ring topology. The results can be seen in Fig. \ref{muchanged}, where the yellow, red, green, and blue (light gray, dark gray, medium gray, and black) areas correspond to $\varepsilon_1=0$, $2$, $4$, and $6$, respectively. 
The interconnection delay is modulated with a sawtooth wave around a mean delay value $\tau_{01}=2\pi$ which is kept constant in both panels. The coupling phase is also fixed at  $\beta_1=0$. The Gershgorin interval  $|\mathrm{Re}(\nu)|\leq\mu$ is bounded by the two solid dashed lines, and the maximum eigenvalue of the adjacency matrix at given $\mu$ is located at the right edge of this interval ($\nu_{\mathrm{max}}= \mu$). Panel (a) corresponds to the gain parameter value $\sigma_1=0.1$. One can see that while the constant delay interconnection ($\varepsilon_1=0$, yellow/light gray region) cannot stabilize the origin at any $\mu$, amplitude death becomes possible by applying time-varying delays as soon as the Gershgorin interval is covered by the stability region. In panel (b) we depict the situation $\sigma_1=0.02$. In this case, a necessary condition for amplitude death is provided by the stability condition (\ref{stabilitycondition}), which in this case reads $\mu>\lambda/(\sigma_1\cos\beta_1)=5$. 
Consequently, amplitude death is achieved for $\mu>5$ and those values of $\varepsilon_1$ for which the Gershgorin interval is contained within the associated stability region.

\subsection{Case II: Self-feedback at each node}

\begin{figure}
\includegraphics[width=\columnwidth,height=!]{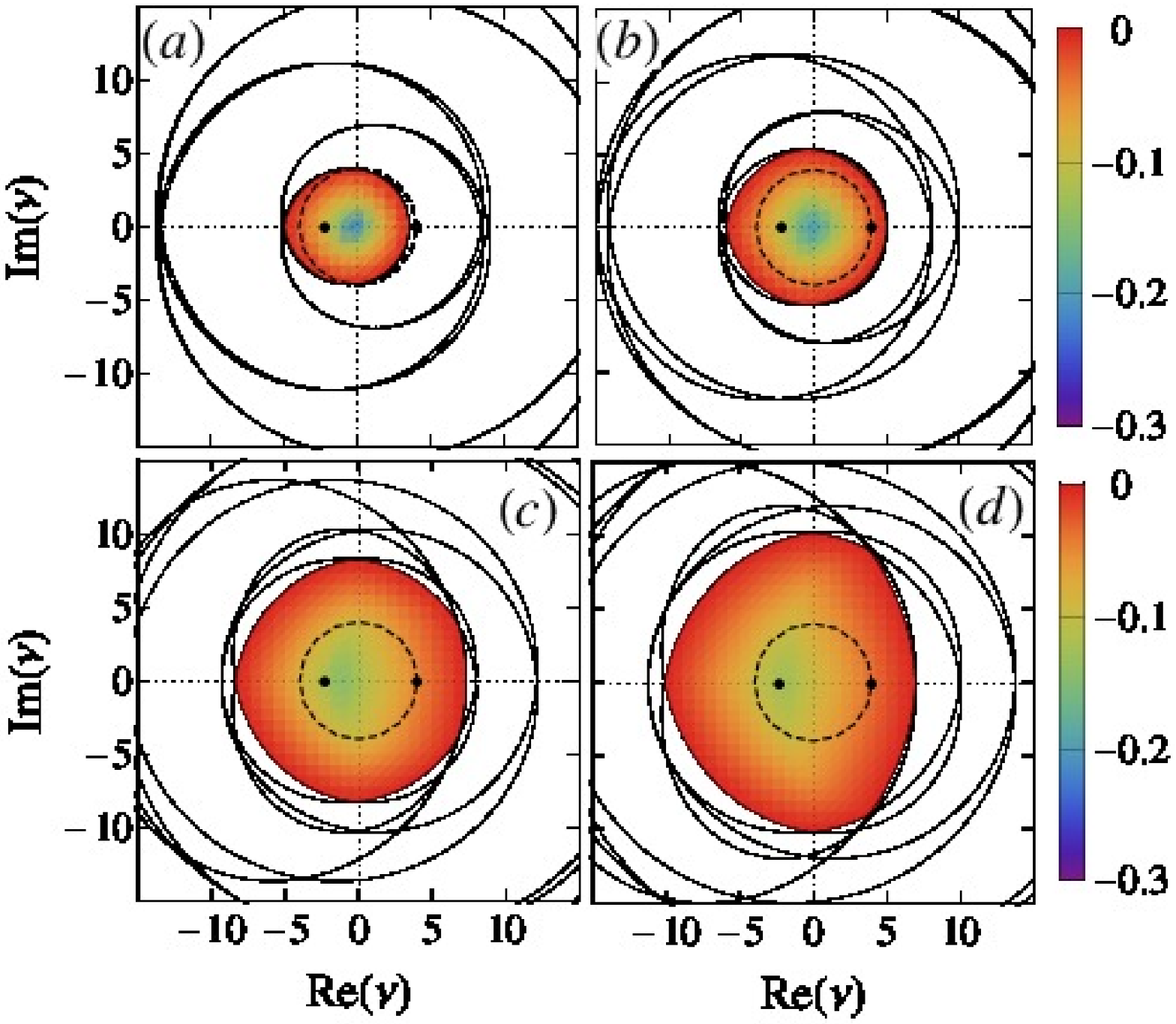}
\caption{(Color online) Master stability function in the complex $\nu$ plane for a regular ring network of Stuart-Landau oscillators with constant delay inter-node connections and a variable-delay self-feedback. The stability region is calculated from Eq. (\ref{charmsfsl}) for a constant interconnection delay $\tau_{01}=2\pi$, with $\sigma_1=0.1$, and a sawtooth-wave modulation of the self-feedback delay in the high-frequency regime around a mean value $\tau_{02}=2\pi$ at a modulation amplitude $\varepsilon_2=2\pi$ and different values of the gain parameter $\sigma_2$:  (a) $\sigma_2=0.05$, (b) $\sigma_2=0.2$, (c) $\sigma_2=0.5$, (d) $\sigma_2=0.7$, and coupling phase $\beta_2=0$. Other parameters as in Fig.~\ref{fig1}. }
%
\label{var1}
\end{figure}

\begin{figure}
\includegraphics[width=\columnwidth,height=!]{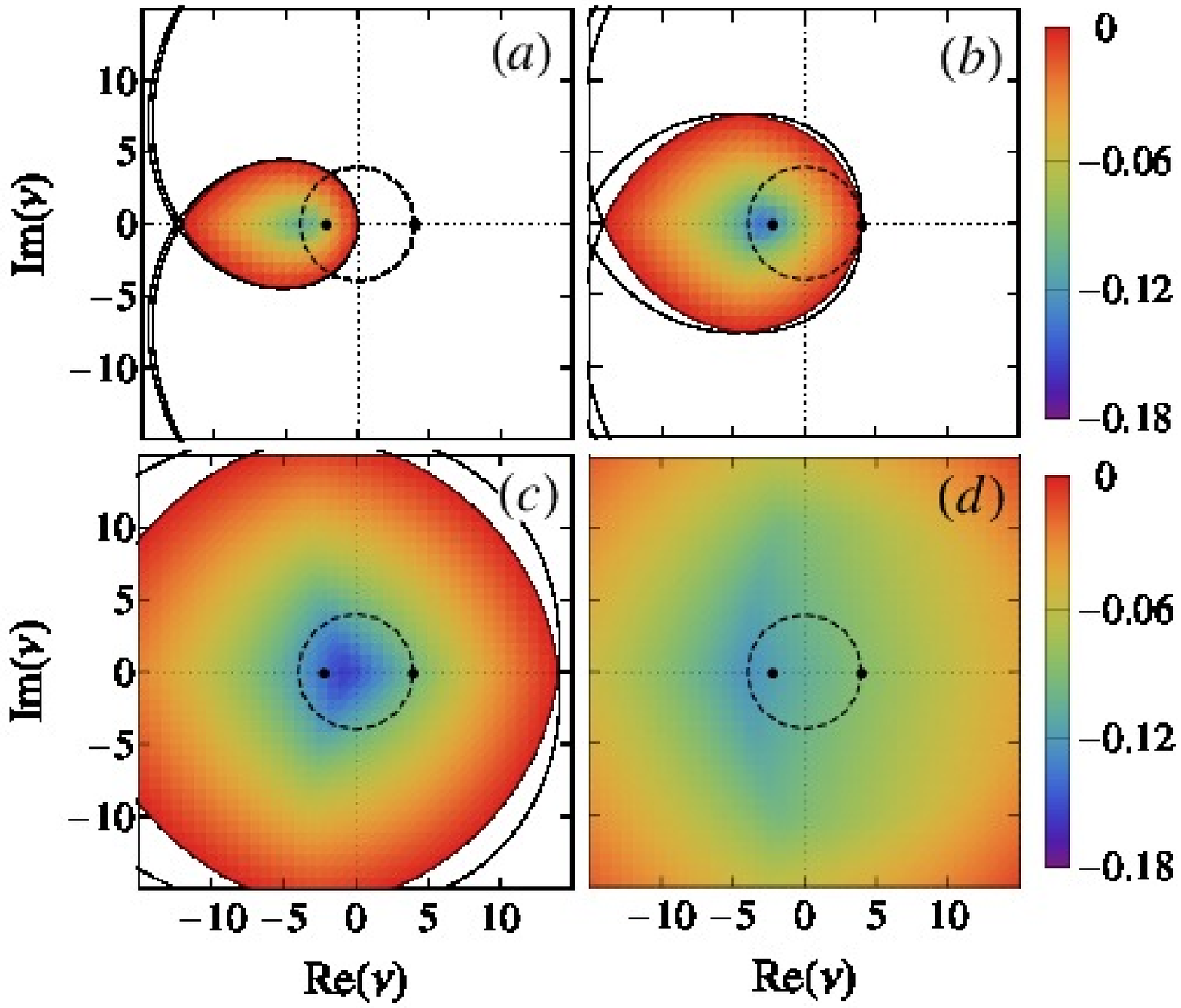}
\caption{(Color online) Master stability function in the complex $\nu$ plane corresponding to Fig. \ref{var1}, with $\sigma_1=0.02$ and different values of the self-feedback gain parameter $\sigma_2$:  (a) $\sigma_2=0.02$, (b) $\sigma_2=0.1$, (c) $\sigma_2=0.3$, (d) $\sigma_2=0.5$.  Other parameters as in Fig. \ref{var1}.}
\label{var2}
\end{figure}

We will now investigate the region of amplitude death in the full system Eqs.~(\ref{sys1})--(\ref{sys2}) which includes self-feedback, by numerically analyzing the corresponding characteristic Eq.~(\ref{charmsfsl}). In Fig. \ref{var1} we show the calculated master stability function for a constant delay inter-node connection and a variable-delay self-feedback with a sawtooth-wave modulation. Throughout this section, we consider a regular ring network topology with $k=2$ interconnectons on each side of a node, and a self-feedback at each node. The adjacency matrix of the interconnection topology has a constant row sum $\mu=4$. The coupling parameters are $\tau_{01}=2\pi$, $\varepsilon_1=0$, $\sigma_1=0.1$ and $\beta_1=0$. The self-feedback parameters are $\tau_{02}=2\pi$, $\varepsilon_2=2\pi$ and $\beta_2=0$, while the gain parameter $\sigma_2$ is different in each panel: (a) $\sigma_2=0.05$, (b) $\sigma_2=0.2$, (c) $\sigma_2=0.5$, (d) $\sigma_2=0.7$. The Gershgorin circle $|\nu|=\mu=4$ is denoted by a dashed line, and the solid black dots are the maximum and minimum eigenvalues of the adjacency matrix of the interconnection topology for $N=20$ oscillators:  $\nu_{\mathrm{min}}\approx-2.236$, $\nu_{\mathrm{max}}=4$. For lower values of the self-feedback gain parameter $\sigma_2$ (panel (a)), the largest eigenvalue of the interconnection adjacency matrix lies outside the stability region, making the origin unstable. By gradually enlarging $\sigma_2$ we observe a monotonic enlargement of the stability region, whose boundaries eventually surpass the Gerschgorin circle (panels (b)--(d)), giving rise to amplitude death above some threshold value of the parameter $\sigma_2$. The enlargment of the stability region with increasing the self-feedback gain parameter is even more pronounced for lower values of the interconnection gain parameter $\sigma_1$. In Fig. \ref{var2} we give the corresponding master stability function for $\sigma_1=0.02$ at various $\sigma_2$.
The other parameters are unchanged with respect to Fig. \ref{var1}. We note that the choice $\sigma_1=0.02$ in the analogous case without self-feedback does not satisfy the stability criterion Eq.~(\ref{stabilitycondition}), thus making amplitude death impossible for any modulation amplitude $\varepsilon_1$ in the inter-node connection delay (see Fig. \ref{fig2}). However, it is seen from Fig.~\ref{var2} that by including self-feedback this restriction is lifted, and amplitude death is achievable at certain parameter values of the self-feedback coupling.   

\begin{figure}
\includegraphics[width=\columnwidth,height=!]{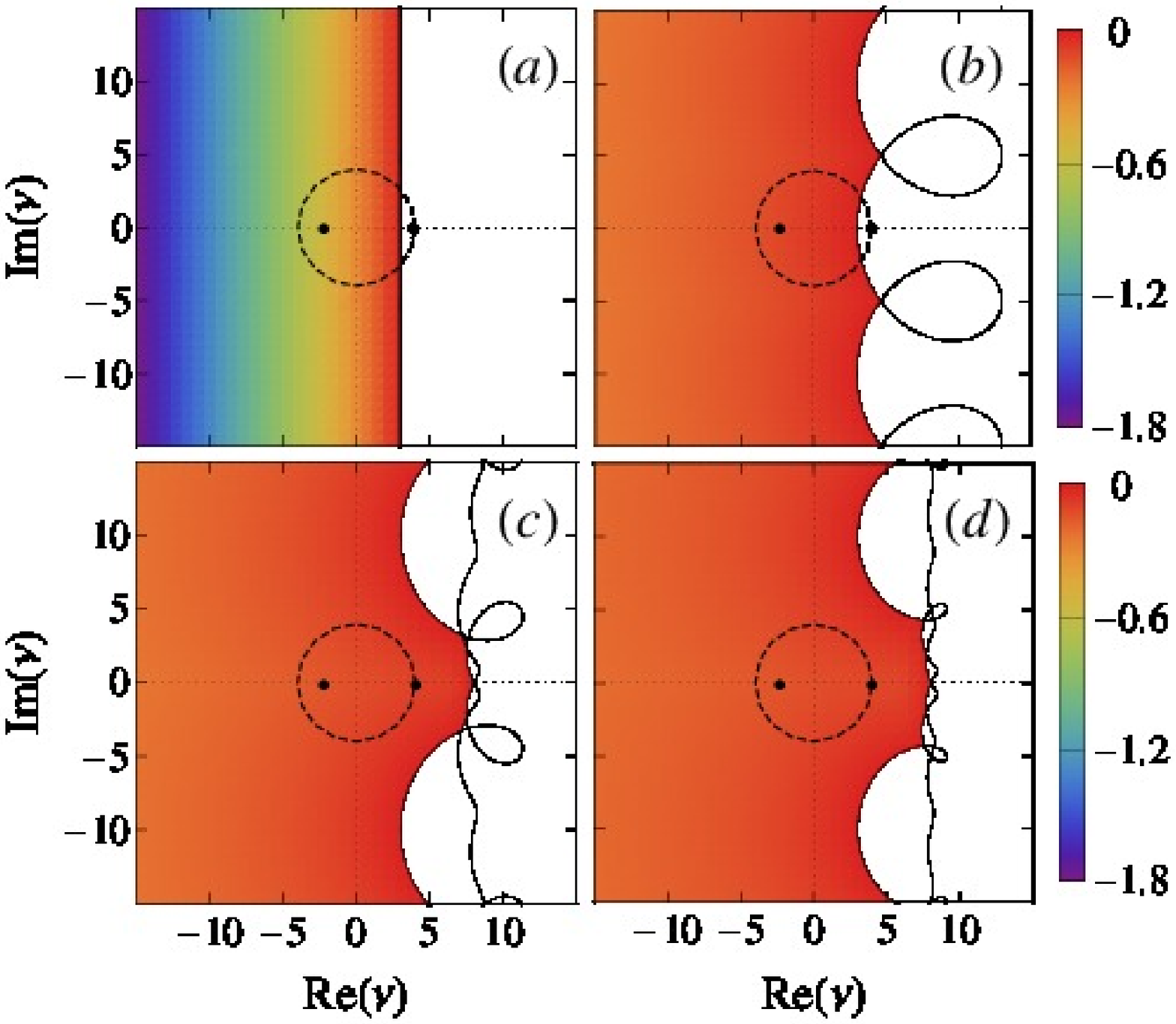}
\caption{(Color online) Master stability function in the $(\mathrm{Re}(\nu),\mathrm{Im}(\nu))$ plane for an instantaneous interaction between oscillators: $\sigma_1=0.1$, $\tau_{01}=0$. (a) case without self-feedback, $\sigma_2=0$. (b) Constant delay in the self-feedback: $\sigma_2=0.5$, $\tau_{02}=2\pi$, $\varepsilon_2=0$. (c) Sawtooth-wave delay modulation in the self-feedback: $\sigma_2=0.5$, $\tau_{02}=2\pi$, $\varepsilon_2=\pi$. (d) Sawtooth-wave delay modulation in the self-feedback: $\sigma_2=0.5$, $\tau_{02}=2\pi$, $\varepsilon_2=2\pi$. Other parameters as in Fig. \ref{var1}. }
\label{var3}
\end{figure}

Previously we have shown that amplitude death cannot be achieved for an instantaneous interacton between oscillators for any regular network topology and any coupling parameters if the self-feedback is absent. In panel (a) of Fig. \ref{var3} we give the master stability function in such case ($\tau_{01}=0$, $\sigma_2=0$) for $\sigma_1=0.1$. In this case, the maximum eigenvalue of the interconnection adjacency matrix lies outside the stability region, rendering the origin unstable. By including a self-feedback, this severe limitation of achieving amplitude death can be overcome. In panels (b)--(d) we show the corresponding stability regions for $\sigma_2=0.5$ for a sawtooth-wave modulation of the self-feedback delay around $\tau_{02}=2\pi$ for increasing modulation amplitude $\varepsilon_2$.
It can be observed that above a certain value of $\varepsilon_2$ (e.g. panels (c) and (d)), the Gershgorin disk is contained in the stability region, enabling amplitude death at these parameter values.

\begin{figure}
\includegraphics[width=\columnwidth,height=!]{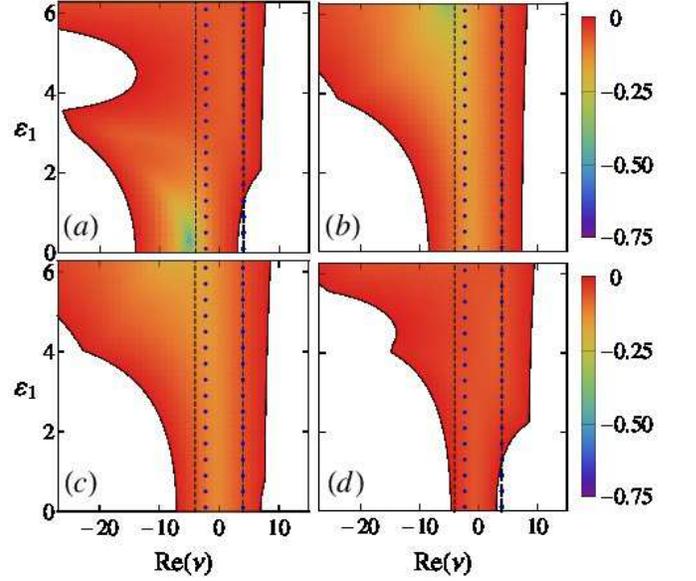}
\caption{(Color online) Master stability function in dependence of the modulation amplitude $\varepsilon_1$ for a sawtooth-wave modulation of the interconnection delay $\tau_1$ and different types of modulatons of the self-feedback delay $\tau_2$: (a) $\varepsilon_2=0$ (constant self-feedback delay); (b) $\varepsilon_2=2\pi$ with a sawtooth-wave modulation; (c) $\varepsilon_2=2\pi$ with a sine-wave modulation; (d) $\varepsilon_2=2\pi$ with a square-wave modulation. The coupling parameters: $\tau_{01}=\tau_{02}=2\pi$, $\sigma_1=0.1$, $\sigma_2=0.5$, $\beta_1=\beta_2=0$. The Gershgorin interval $[-4, 4]$ is contained between the black dashed lines, and the blue dotted lines correspond to the maximum and minimum eigenvalues of the interconnection adjacency matrix. Other parameters as in Fig. \ref{var1}.}
\label{var4}
\end{figure}

\begin{figure}
\includegraphics[width=\columnwidth,height=!]{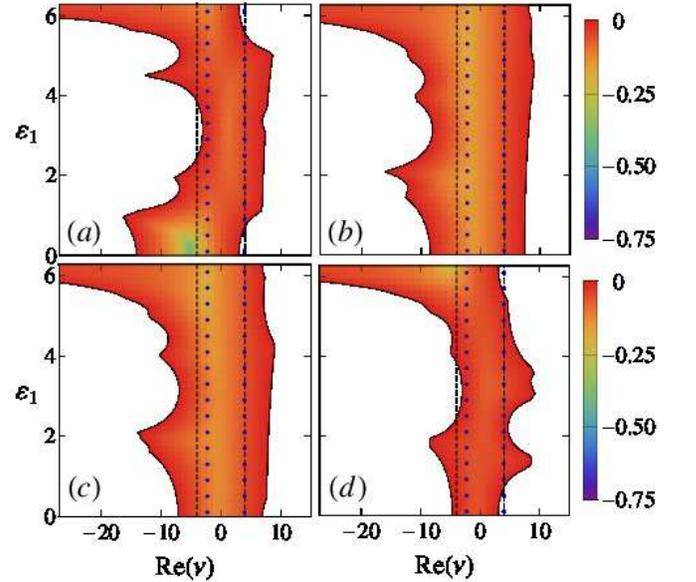}
\caption{(Color online) Master stability function corresponding to Fig. \ref{var4} for a square-wave modulation of the interconnection delay $\tau_1$. Other parameters as in Fig. \ref{var4}.}
\label{var5}
\end{figure}

The enhancement of the stability region by including variable-delay self-feedback is also observed when the delay interaction between the oscillators is time-varying. In Fig. \ref{var4} we depict the master stability function in the parametric plane of the modulation amplitude $\varepsilon_1$ of the inter-node connection delay and the real part of the eigenvalue $\nu$ of the interconnection adjacency matrix. In each panel, the modulation of the interconnection delay is with a sawtooth wave around a mean value $\tau_{01}=2\pi$, and the fixed system parameters are the self-feedback mean delay $\tau_{02}=2\pi$, the gain values $\sigma_1=0.1$ and $\sigma_2=0.5$, and the coupling phases $\beta_1=\beta_2=0$. Panel (a) shows the stability region for a constant delay in the self-feedback ($\varepsilon_2=0$). Comparing to the case without self-feedback in panel (d) of Fig. \ref{fig3}, it is observed that although the stability region is expanded considerably towards the negative values of $\mathrm{Re}(\nu)$, the 
range of $\varepsilon_1$ at which amplitude death is achieved stays almost unchanged. Namely, at low values of $\varepsilon_1$, in the approximate interval $\varepsilon_1\in[0,1.2]$, the maximum eigenvalue of the interconnection adjacency matrix for the considered ring topology is outside the stability region if the self-feedback is absent or with a constant delay.  As the delay in the self-feedback is modulated, the stability region is changing depending on the type of the delay modulation. If the self-feedback delay $\tau_2$ is modulated with a sawtooth-wave (panel (b)) or a sine wave (panel (c)), the stability domain expands monotonically, until it completely covers the Gershgorin's interval $[-4,4]$, making the amplitude death possible for every value of $\varepsilon_1$. The value of the delay modulation amplitude in the self-feedback in panels (b) and (c) is $\varepsilon_2=2\pi$. In the case of a square-wave modulation of $\tau_2$, the expansion is non-monotonic, and for the same modulation amplitude $\varepsilon_2=2\pi$ (panel (d)), the instability interval for $\varepsilon_1$ is unchanged with respect to the case in panel (a) for a constant self-feedback delay. The positive influence of the self-feedback on increasing the stability region is also observed for other types of modulations of the interconnection delay, and in Fig. \ref{var5} we show the master stability function for a square-wave modulation corresponding to Fig. \ref{var4}. From the resulting stability diagrams, it can be concluded that the parameter intervals for amplitude death are expanded considerably by including a variable-delay self-feedback, and the positive effects of such an inclusion depend on the type of the delay modulation. 

\begin{figure}
\includegraphics[width=\columnwidth,height=!]{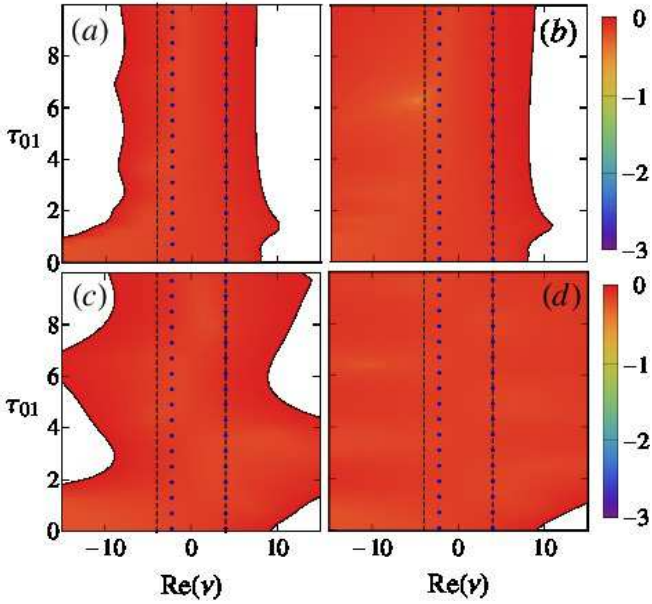}
\caption{(Color online) Master stability function in the $(\tau_{01},\mathrm{Re}(\nu))$ plane. The delay modulation in the self-feedback is in a form of a sawtooth-wave around a mean value $\tau_{02}=2\pi$ with amplitude $\varepsilon_2=2\pi$. (a) Constant inter-node connection delay, parameters: $\sigma_1=0.1$, $\sigma_2=0.5$, $\varepsilon_1=0$; (b) Sawtooth-wave modulation of the interconnection delay, parameters: $\sigma_1=0.1$, $\sigma_2=0.5$, $\varepsilon_1=\tau_{01}$;  (c) Constant interconnection delay, parameters: $\sigma_1=0.02$, $\sigma_2=0.2$, $\varepsilon_1=0$; (d) Sawtooth-wave modulation of the interconnection delay, parameters: $\sigma_1=0.02$, $\sigma_2=0.2$, $\varepsilon_1=\tau_{01}$. Other parameters as in Fig. \ref{var1}.}
\label{var6}
\end{figure}

Application of a delayed self-feedback in the ring oscillator network can also expand the interval of the interconnection mean delay $\tau_{01}$ leading to amplitude death. In Fig. \ref{var6} we have numerically calculated the master stability function in the $(\tau_{01},\mathrm{Re}(\nu))$ plane for a sawtooth-wave modulation of the delay in the self-feedback around a mean value $\tau_{02}=2\pi$ with modulation amplitude $\varepsilon_2=2\pi$. The Gershgorin interval is marked by the black dashed lines, and the eigenspectrum of the interconnection adjacency matrix is contained within the interval bounded by the blue dotted lines. In panel (a) we show the stability region for $\sigma_1=0.1$ and $\sigma_2=0.5$ for a constant interconnection delay ($\varepsilon_1=0$). Comparing to the case without self-feedback (panel (a) in Fig. \ref{fig4}), it is seen that the Gershgorin interval is completely contained in the stability region if the self-feedback is included, and amplitude death now occurs at each value of $\tau_{01}$ within the interval depicted in the panel. The stability region becomes even larger if the interconnection delay is also modulated, and in panel (b) we show the case of a sawtooth-wave modulated inter-node connection delay $\tau_1$ at a maximum possible amplitude $\varepsilon_1=\tau_{01}$. 
In panels (c) and (d) we depict the master stability function corresponding to panels (a) and (b), respectively, for $\sigma_1=0.02$ and $\sigma_2=0.2$ and other parameters unchanged. If the self-feedback were not present, the stability region would be given by the corresponding panels (c) and (d) in Fig. \ref{fig4}, in which case the system parameters are such that the stability condition Eq.~(\ref{stabilitycondition}) is not satisfied, and amplitude death becomes impossible at any delay value. By introducing variable-delay self-feedback, the stability region expands for both constant delay inter-node connection (panel (c) in Fig. \ref{var6}) and a variable-delay inter-node connection (panel (d) in Fig. \ref{var6}), enabling amplitude death for the whole depicted range of $\tau_{01}$.

\begin{figure}
\includegraphics[width=\columnwidth,height=!]{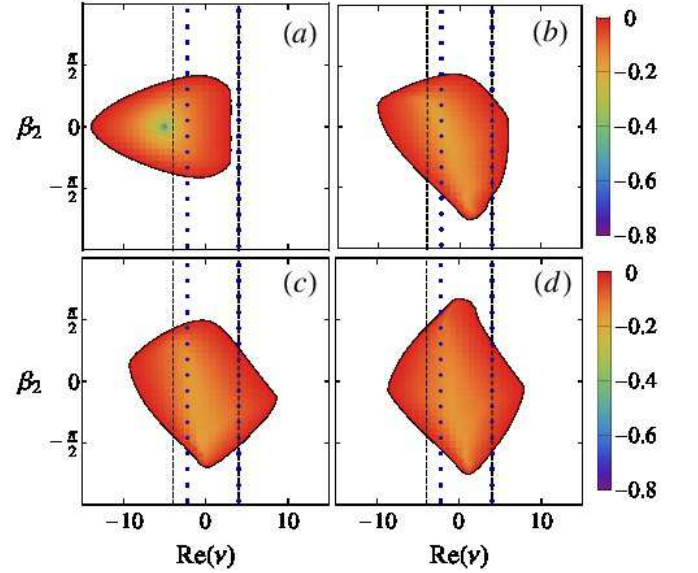}
\caption{(Color online) Master stability function in the $(\beta_2,\mathrm{Re}(\nu))$ plane for a constant delay interconnection $\tau_1=2\pi$ and sawtooth-wave modulation of the self-feedback delay around a mean value $\tau_{02}=2\pi$ and different modulation amplitudes: (a) $\varepsilon_2=0$, (b) $\varepsilon_2=2\pi/3$, (c) $\varepsilon_2=4\pi/3$, (d) $\varepsilon_2=2\pi$. The values of the gain parameters are $\sigma_1=0.1$ and $\sigma_2=0.5$. Other parameters as in Fig. \ref{var1}.}
\label{var7}
\end{figure}

The influence of the coupling phase $\beta_2$ on amplitude death can be seen by calculating the master stability function from the characteristic Eq. (\ref{charmsfsl}) in the plane spanned by $\beta_2$ and $\mathrm{Re}(\nu)$. The resulting stability region is depicted in Fig. \ref{var7}. Different panels correspond to different modulation amplitudes of the self-feedback delay $\tau_2$: (a) $\varepsilon_2=0$ (constant delay), (b) $\varepsilon_2=2\pi/3$, (c) $\varepsilon_2=4\pi/3$, (d) $\varepsilon_2=2\pi$. The interconnection delay $\tau_1$ is taken constant and equal to its mean value $\tau_1=\tau_{01}=2\pi$, and the self-feedback delay $\tau_2$ is modulated with a sawtooth-wave around $\tau_{02}=2\pi$. The gain parameters are set to $\sigma_1=0.1$ and $\sigma_2=0.5$, and the coupling phase in the interconnection is set to zero ($\beta_1=0$). For a constant self-feedback delay (panel (a)), the rightmost value of the Gershgorin interval, i.e., the maximum eigenvalue of the interconnection adjacency matrix, is outside the stability region for any $\beta_2$, making amplitude death impossible in this non-modulated case. For increasing modulation amplitude in the self-feedback delay (panels (b)--(d)), the stability region expands, eventually covering the whole Gershgorin interval in a certain finite range of $\beta_2$, in which amplitude death is achieved. 

\subsection{Case III: Self-feedback at a single node}

We now investigate the possibility of inducing amplitude death in a regular ring network topology of Stuart-Landau oscillators by applying a variable-delay self-feedback at a single node only. The system dynamics is now governed by:
\begin{align}
\dot z_j=h(z_j)+\sigma_1e^{i\beta_1}\sum_{n=1}^{N}a_{jn}\left[z_n(t-\tau_1(t))-z_j(t)\right]\nonumber\\
+\delta_{1j}\,\sigma_2e^{i\beta_2}\left[z_j(t-\tau_2(t))-z_j(t)\right]
\label{sys1a}
\end{align}
with $j=1,2\dots N$, where the local dynamics is given by Eq.~(\ref{sys2}). This system differs essentially from the system Eq.~(\ref{sys1}), in which the self-feedback was applied at each node, by the presence of the Kronecker delta $\delta_{1j}$ in the rightmost term, indicating a self-feedback at the first node only. Since the resulting system cannot be treated via the master stability formalism, we will investigate the network dynamics directly, and determine the parameters leading to amplitude death by numerically analyzing the system Eq.~(\ref{sys1a}). For that purpose, we integrate the system and follow the time evolution of the dynamical variables $x_j$ and $y_j$.  

\begin{figure*}
\includegraphics[width=\textwidth,height=!]{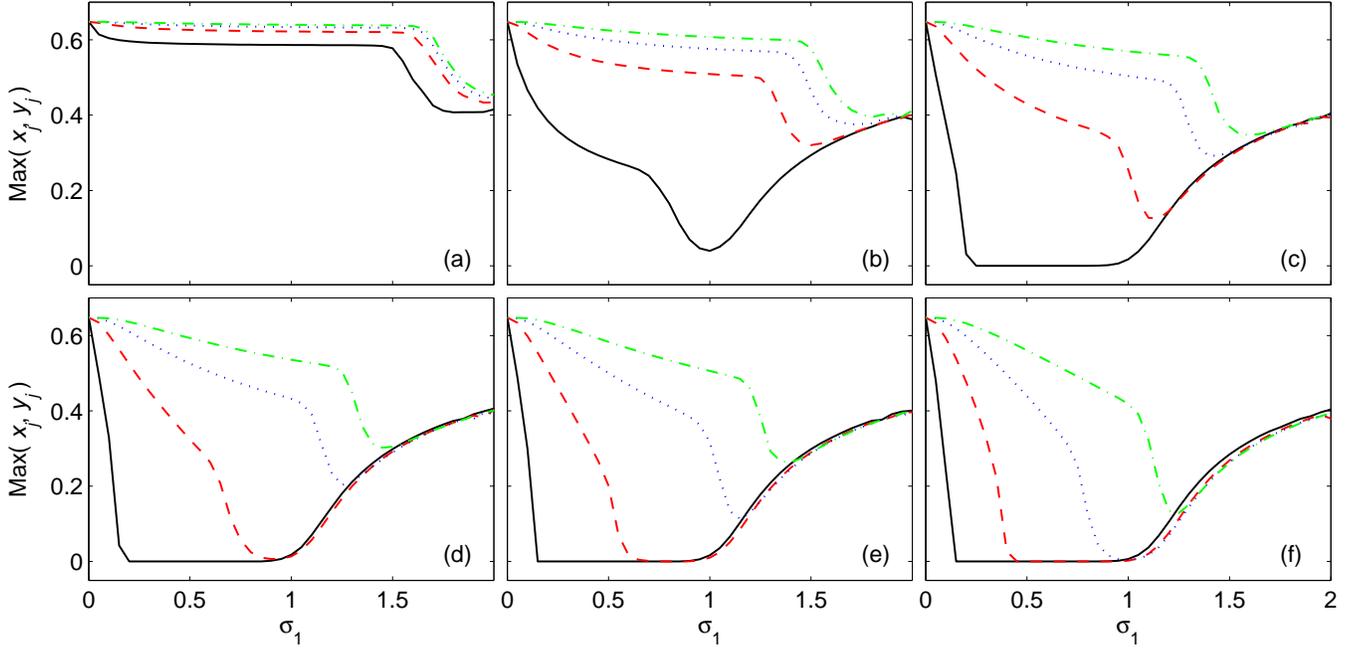}
\caption{(Color online) Maximum amplitude of the oscillations of a regular ring network of coupled Stuart-Landau oscillators as a function of the strength $\sigma_1$ of the inter-node connection. The inter-node coupling parameters are $\tau_{01}=2\pi$, $\varepsilon_1=0$ and $\beta_1=0$. The self-feedback acts at a single node in form of a sawtooth-wave with $\tau_{02}=2\pi$, $\varepsilon_2=2\pi$, $\varpi_2=10$, and $\beta_2=0$. Self-feedback gain:  (a) $\sigma_2=0.1$, (b) $\sigma_2=0.5$, (c) $\sigma_3=1$, (d) $\sigma_2=1.5$, (e) $\sigma_2=2$, (f) $\sigma_2=4$. The solid black, dashed red, dotted blue, and dash-dotted green lines correspond to increasing number of nodes in the oscillator network: $N=5, 10, 15,$ and 20, respectively. Other parameters as in Fig.~\ref{fig1}.
}
\label{sn1}
\end{figure*}

In panels (a)--(f) of Fig. \ref{sn1} we summarize the results of the numerical simulations by depicting the dependence of the maximum amplitude of the system variables $x_j$ and $y_j$ on the gain parameter $\sigma_1$ representing the strength of the inter-node connection. The maximum amplitude is calculated from a sample of all $2N$ variables $x_j$ and $y_j$ taken after a long transient from randomly chosen initial conditions in the interval $[0,1]$.  We choose a constant inter-node connection delay $\tau_{01}=2\pi$, and a sawtooth-wave modulation of the self-feedback delay $\tau_2$ around a mean value $\tau_{02}=2\pi$ with amplitude $\varepsilon_2=2\pi$ and a frequency $\varpi_2=10$. The coupling phases are fixed at $\beta_1=\beta_2=0$. The parameters of the local dynamics are $\lambda = 0.1$, $\omega = 1$, and $\gamma = 0.1$, as before. Different panels correspond to different values of the self-feedback gain: (a) $\sigma_2=0.1$, (b) $\sigma_2=0.5$, (c) $\sigma_3=1$, (d) $\sigma_2=1.5$, (e) $\sigma_2=2$, (f) $\sigma_2=4$. Each panel contains four different resulting curves denoted by solid black, dashed red, dotted blue, and dash-dotted green lines corresponding to different number of nodes in the oscillator network ($N=5, 10, 15,$ and 20, respectively). In each case, the network topology is a ring with $k=2$ interconnections at each side of a node. The intervals of $\sigma_1$ that lead to amplitude death are indicated by the diminishing maximum amplitude visualized by the horizontal plateau at zero amplitude. Although for low values of the self-feedback strength $\sigma_2$ [panels (a) and (b)] amplitude death cannot be achieved, the stabilization of the origin becomes possible as $\sigma_2$ is increased above a certain threshold value [panels (c)--(f)]. In the latter cases, it is observed that the $\sigma_1$ interval for amplitude death strongly depends on the number of nodes in the network, becoming wider as the number of nodes $N$ is decreased. As the number of nodes is increased, the amplitude death interval becomes narrower, eventually dissapearing at higher values of $N$. 

\begin{figure}
\includegraphics[width=\columnwidth,height=!]{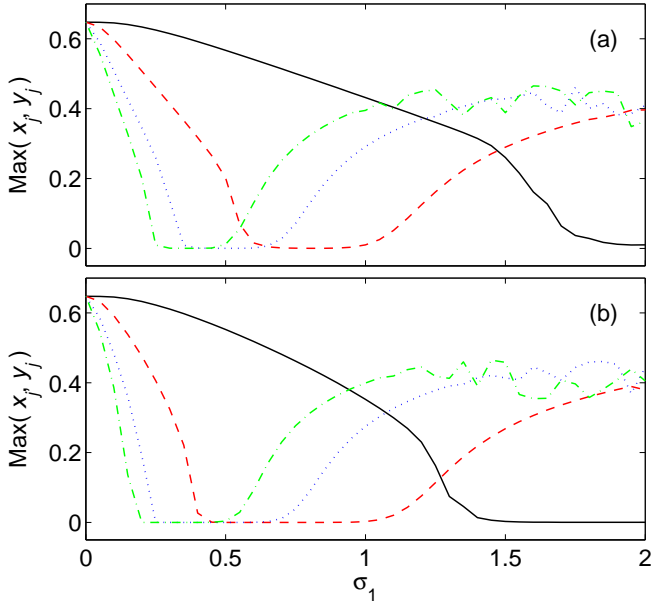}
\caption{(Color online) Maximum amplitude of the oscillations as a function of $\sigma_1$ for a regular ring network of $N=10$ coupled Stuart-Landau oscillators with varying number of interconnections $k$ on each side of a node: $k=1$ (solid black), $k=2$ (dashed red), $k=3$ (dotted blue), $k=4$ (dash-dotted green). Self-feedback gain parameter: (a) $\sigma_2=2$, (b) $\sigma_2=4$. Other parameters as in Fig. \ref{sn1}.}
\label{sn2}
\end{figure}

Figure \ref{sn2} depicts the dependence of the amplitude death interval on the node degree in the interconnection topology. Panel (a) correspond to $\sigma_2=2$ and panel (b) to $\sigma_2=4$. The number of nodes at each panel is $N=10$, and solid black, dashed red, dotted blue, and dash-dotted green curves correspond to increasing number of interconnections at each side of a node ($k=1, 2, 3,$ and 4, respectively). The other parameters are unchanged with respect to Fig. \ref{sn1}. It is observed that as the number of connections between the nodes is increased, the amplitude death interval becomes narrower, being shifted towards smaller values of $\sigma_1$.  

\begin{figure}
\includegraphics[width=\columnwidth,height=!]{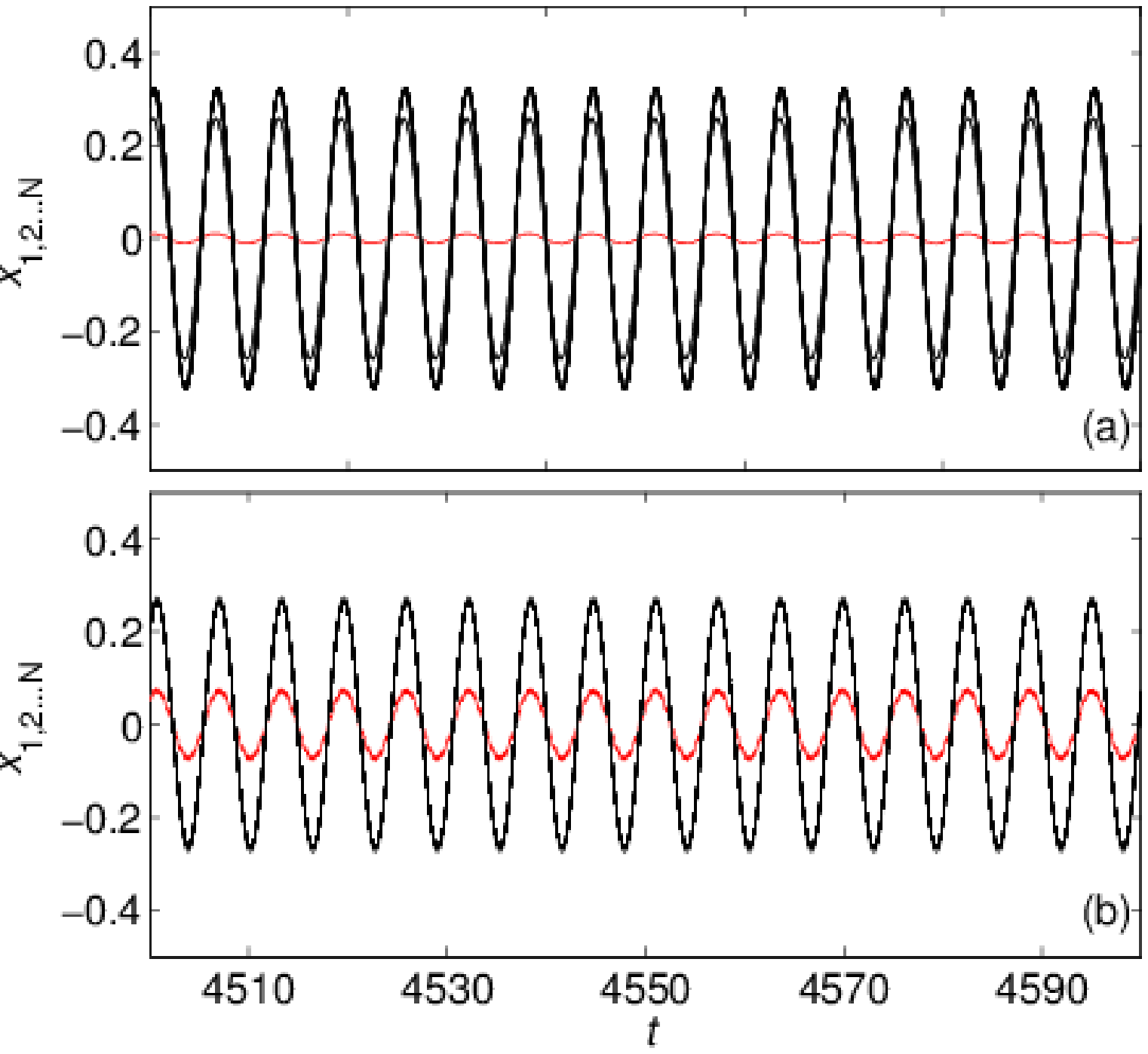}
\caption{(Color online) Partial amplitude death: Time series $x_1(t)$ (red) and $x_{2...N}(t)$ (black) for $\sigma_1=0.05$, $\sigma_2=2.5$ and $N=20$. Number of interconnections at each side of a node: (a) $k=1$, (b) $k=9$. Other parameters as in Fig. \ref{sn1}.}
\label{sn3}
\end{figure}

It is interesting to note that for small enough values of the inter-node connection strength $\sigma_1$, large value of the self-feedback gain $\sigma_2$, and a large number of nodes $N$, the oscillator at which self-feedback is applied performs a small-amplitude oscillation around the origin, while the dynamics of the rest of the oscillators are almost unaffected by the self-feedback, and they continue to oscillate at large amplitude exhibiting phase synchronization. This regime of partial amplitude death is depicted in Fig. \ref{sn3} for $\sigma_1=0.05$, $\sigma_2=2.5$ and $N=20$. Panel (a) depicts the time series after a long transient for a regular ring network with one interconnection at each side of a node ($k=1$). The dynamics of the oscillator with self-feedback is given by the solid red (gray) curve, and the time series of the other oscillators are depicted by solid black curves. As the coupling range $k$ is increased, the amplitude of the oscillations of the node with the self-feedback increases, and in panel (b) we show the associated dynamics for $k=9$ interconnections at each side. Accordingly, the phase synchronization of the rest of the nodes in panel (a) turns into complete (amplitude and phase) synchronization in panel (b). 

\section{Conclusions}

We have shown that amplitude death can be induced by applying coupling with a time-varying delay in a network of oscillators even if it does not exist in the case of constant delay, and its regime in the parameter space can be enhanced significantly. In the case of a regular network topology with delayed coupling, which may include a Pyragas-type self-feedback at each node, a master stability function formalism has been employed to analyze the linear stability of the unstable homogeneous steady state, and predict analytically the boundaries of stability. At high-frequency delay modulation, analytical results for the occurrence of amplitude death can be obtained by approximating the variable-delay coupling terms by distributed-delay with delay distribution kernels matching the probability density function. The success of the proposed method has been demonstrated both numerically and analytically for a regular ring network consisting of Stuart-Landau limit cycle oscillators in the regime near a Hopf 
bifurcation. We have shown that controllability of the network fixed point is strongly limited by the local node dynamics, which could be removed in certain cases by including a variable-delay self-feedback. In addition, we have shown that amplitude death can even be induced if the self-feedback is applied at a single node only, for certain control parameter intervals and not too large networks. 

With respect to the practical realization of amplitude death in real systems using the proposed variable-delay coupling methods, it must be emphasized that for the considered high-frequency regime of the delay modulations, the distributed-delay approximation of the network dynamics does not depend on the frequency of the modulations. This allows for considerable flexibility in the choice of the delay modulation, i.e., since any modulation which corresponds to the same probability density function of the delay distribution will lead to the same stabilization regimes of the steady state. Specifically, choosing $\tau_i(t)$ in all inter-node connections with different frequencies and initial phases, keeping the same mean delay $\tau_{01}$ and modulation amplitude $\varepsilon_1$ will still result in stabilization. 
It has been demonstrated that the high-frequency condition is not very severe in real applications, and the distributed-delay limit is still valid for fairly low frequencies, making the distributed-delay approximation of the variable-delay systems a versatile method of analysis for such systems \cite{JUE12}.

The variability of the coupling delay in real networks is often due to random fluctuations induced by the environment, or due to imperfections of the system. In these cases, the delay varies stochastically in time, being distributed over an interval of values and characterized by a distribution function. This situation is also covered by our analysis, since fast random fluctuations of the delay are equivalent to a deterministic modulation with a delay distribution $\rho$ equal to that of the random case. Consequently, random delay fluctuations with a constant probability distribution in a certain interval are equivalent to a deterministic variation of the delay with a sawtooth-wave modulation, independently of the skewness of the sawtooth-wave, and fast random fluctuations between two discrete delays is equivalent to a square-wave modulation, etc. Hence, by invoking noise in the delay lines artificially, one may enhance the regions of amplitude death. In addition, by using digital variable-delay lines with deterministic or stochastic delay variations, one may in principle achieve any desired form of the delay distribution.

\begin{acknowledgments}
This work was supported by Deutsche Forschungsgemeinschaft in the framework of SFB 910: ``Control of self-organizing
nonlinear systems: Theoretical methods and concepts of application.'' 
\end{acknowledgments}

\begin{appendix}

\section{Eigenspectrum of the adjacency matrix for ring network topologies}

\begin{figure}
\includegraphics[width=\columnwidth,height=!]{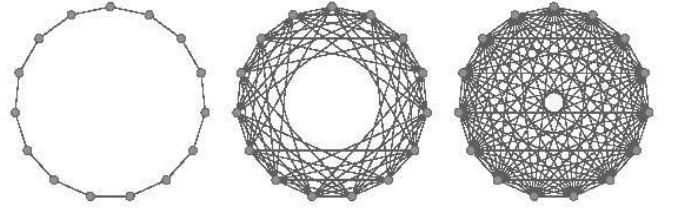}
\caption{Regular $2k$-ring network consisting of 15 nodes. Each node is connected to $2k$ nearest neighbors, with $k$ links on each side of the node: (left) $k=1$, bidirectional ring; (center) $k=5$; (right) $k=7$ complete (all-to-all) coupling.}
\label{smallworld}
\end{figure}

The stability of the collective fixed point $\mathbf{X}^*$ at the origin is determined by the position of the eigenvalues $\nu_m$ of the adjacency matrix $\widehat{\mathbf{A}}$. The fixed point of the network dynamics is stable if all the eigenvalues $\nu_m$ are located inside the stability region determined by the master stability function. We consider a regular network with a ring topology, consisting of $N$ nodes, where each node is bidirectionally coupled to $2k$ nearest neighbors with $k$ links (edges) on each side of the node (see Fig. \ref{smallworld}). 
The degree of each node is thus constant and equal to $2k$. The adjacency matrix $\widehat{\mathbf{A}}_k$ for this network topology is an $N\times N$ symmetric circulant matrix with constant row sum $\mu=2k$. For example, the adjacency matrix for a bidirectional ring network with a single link between each two adjacent nodes ($k=1$) is    
\begin{equation}
\widehat{\mathbf{A}}_{k=1}=
\left(
\begin{array}{ccccccc}
0 & 1 & 0 & 0 & \cdots & 0 & 1  \\
1 & 0 & 1 & 0 & \cdots & 0 & 0  \\
0 & 1 & 0 & 1 & \cdots & 0 & 0  \\
\vdots & \vdots & \vdots & \vdots & \ddots & \vdots & \vdots  \\
0 & 0 & 0 & 0 & \cdots & 0 & 1  \\
1 & 0 & 0 & 0 & \cdots & 1 & 0  \\
\end{array}
\right)
\end{equation} 
The adjacency matrix of this $2k$-ring topology can be conveniently represented as a sum of terms involving powers of the elementary $N\times N$ circulant matrix $\widehat{\mathbf{E}}$ \cite{MIE12}:
\begin{equation}
\widehat{\mathbf{E}}=
\left(
\begin{array}{ccccccc}
0 & 0 & 0 & 0 & \cdots & 0 & 1  \\
1 & 0 & 0 & 0 & \cdots & 0 & 0  \\
0 & 1 & 0 & 0 & \cdots & 0 & 0  \\
\vdots & \vdots & \vdots & \vdots & \ddots & \vdots & \vdots  \\
0 & 0 & 0 & 0 & \cdots & 0 & 0  \\
0 & 0 & 0 & 0 & \cdots & 1 & 0  \\
\end{array}
\right)
\end{equation} 
The eigenvalues $E_m$ of $\widehat{\mathbf{E}}$ are easily obtained, since
\begin{equation}
\widehat{\mathbf{E}}
\left(
\begin{array}{c}
x_1\\
x_2\\
x_3\\
\vdots\\
x_{N-1}\\
x_N\\
\end{array}
\right)=
\left(
\begin{array}{c}
x_N\\
x_1\\
x_2\\
\vdots\\
x_{N-2}\\
x_{N-1}\\
\end{array}
\right)
\end{equation}
From the eigenvalue equation $\widehat{\mathbf{E}}\mathbf{x}=E\mathbf{x}$, we obtain the system
$Ex_1=x_N$, $Ex_2=x_1$, \dots ,$Ex_N=x_{N-1}$, which after subsequent multiplicaton leads to $\prod_{n=1}^N x_n=E^N\prod_{n=1}^N x_n$, from which $E^N=1$, and thus the eigenvalue spectrum of $\widehat{\mathbf{E}}$
\begin{equation}
E_m=e^{2\pi i (m-1)/N},
\label{uniring}
\end{equation}
where $m=1,2,\dots,N$.
Incidentally, this is the eigenvalue spectrum of a \textit{unidirectional ring network}, since the elementary circulant matrix $\widehat{\mathbf{E}}$ coincides with the adjacency matrix of such a network.  
$k$-fold application of $\widehat{\mathbf{E}}$ has the effect of $k$ downshifts of the elements of each column vector, and leads to the resulting matrix $\widehat{\mathbf{E}}^k$. The inverse matrix $\widehat{\mathbf{E}}^{-1}$ is 
\begin{equation}
\widehat{\mathbf{E}}^{-1}=
\left(
\begin{array}{ccccccc}
0 & 1 & 0 & 0 & \cdots & 0 & 0  \\
0 & 0 & 1 & 0 & \cdots & 0 & 0  \\
0 & 0 & 0 & 1 & \cdots & 0 & 0  \\
\vdots & \vdots & \vdots & \vdots & \ddots & \vdots & \vdots  \\
0 & 0 & 0 & 0 & \cdots & 0 & 1  \\
1 & 0 & 0 & 0 & \cdots & 0 & 0  \\
\end{array}
\right)
\end{equation} 
The adjacency matrix $\widehat{\mathbf{A}}_k$ of a $2k$-ring network topology can be written in terms of the powers of $\widehat{\mathbf{E}}$ as:
\begin{equation}
 \widehat{\mathbf{A}}_k=\sum_{j=1}^k \left(\widehat{\mathbf{E}}^j+\widehat{\mathbf{E}}^{-j}\right).
\end{equation}
The eigenvalues $\nu_m$ of the adjacency matrix $\widehat{\mathbf{A}}_k$ can be found from rewriting $\widehat{\mathbf{A}}_k\mathbf{x}=\nu\mathbf{x}$ as:
\begin{equation}
\sum_{j=1}^k \left(\widehat{\mathbf{E}}^j+\widehat{\mathbf{E}}^{-j}\right)\mathbf{x}=\nu\mathbf{x}
\end{equation}
from which we obtain
\begin{equation}
\sum_{j=1}^k \left({E_m}^j+{E_m}^{-j}\right)\mathbf{x}=\nu_m\mathbf{x}.
\end{equation}
By taking into account Eq. (\ref{uniring}), we obtain
\begin{equation}
\nu_m=\sum_{j=1}^k \left[\left(e^{\frac{2\pi i}{N}(m-1)}\right)^j+\left(e^{-\frac{2\pi i}{N}(m-1)}\right)^j\right],
\end{equation}
which can be further simplified to 
\begin{equation}
\nu_m=\frac{\displaystyle\sin\left[\frac{(2k+1)\pi (m-1)}{N}\right]}{\displaystyle\sin\left(\frac{\pi (m-1)}{N}\right)}-1,
\label{eigenvalues}
\end{equation}
where $m=1,2,\dots,N$.
The eigenvalues $\nu_m$ are real since $\widehat{\mathbf{A}}_k$ is symmetric, and the corresponding eigenvectors $\mathbf{u}_m$ form an orthogonal basis. Also, the largest eigenvalue of the adjacency matrix in a regular network equals the node degree, which in this case follows from Eq. (\ref{eigenvalues}) for $m=1$, that is $\nu_1=2k$. The corresponding eigenvector $\mathbf{u}_1=(1,1,\dots,1)^T$ is an $N$-dimensional column vector with all entries one. 

From Eq. (\ref{eigenvalues}) one can explicitly derive the formulas for the eigenvalue spectrum for some special cases of regular ring-network topologies, such as \textit{bidirectional ring network} ($k=1$)
\begin{equation}
\nu_m=2\cos\left[\frac{2\pi (m-1)}{N}\right], 
\label{biring}
\end{equation}
and \textit{all-to-all coupling}  ($2k+1=N$), i.e., a \textit{complete graph},
\begin{equation}
\nu_m=
\begin{cases}
N-1, & m=1,\\
-1, & m=2,3,\dots,N.
\end{cases}
\label{alltoall}
\end{equation}

\end{appendix}

\bibliographystyle{prsty-fullauthor}

\end{document}